\documentclass[runningheads]{llncs}
\pdfoutput=1
\usepackage{calc}
\usepackage{tikz}
\usetikzlibrary{decorations.pathreplacing}
\usetikzlibrary{decorations.markings}
\tikzstyle{vertex}=[circle, draw, inner sep=0pt, minimum size=3pt]
\newcommand{\vertex}{\node[vertex]}

\usepackage{graphicx}
\usepackage{amsmath}
\usepackage{amsfonts}
\usepackage{xcolor}
\usepackage{amssymb}
\usepackage{graphicx}
\usepackage{tikz}
\usetikzlibrary{circuits.logic.US}
\usetikzlibrary{positioning}
\usetikzlibrary{matrix}
\usepackage{graphicx}
\usepackage{amsmath}
\newcommand{\boundellipse}[3]
{(#1) ellipse (#2 and #3)
}
\definecolor{magenta}{rgb}{0.8, 0.0, 0.8}
\definecolor{cyan}{rgb}{0.0, 1.0, 1.0}
\definecolor{green1}{rgb}{0.1, 0.6, 0.01}
\definecolor{green}{rgb}{0.11, 0.35, 0.02}
\definecolor{brown}{rgb}{0.65, 0.16, 0.16}
\definecolor{cadetgrey}{rgb}{0.57, 0.64, 0.69}

\newtheorem{observation}{\bf Observation}

\title{Globally Minimal Defensive Alliances: A Parameterized Perspective}
\author{Ajinkya Gaikwad \and   Soumen Maity  }
\authorrunning{ A.\,Gaikwad and S.\,Maity}
\institute{Indian Institute of Science Education and Research, Pune, India 
\email{\texttt{ajinkya.gaikwad@students.iiserpune.ac.in;}}
\email{\texttt{soumen@iiserpune.ac.in}}
}
\begin{document}

\maketitle

\begin{abstract}
A defensive alliance in an undirected graph $G=(V,E)$ is a non-empty set of vertices $S$ satisfying the 
condition that every vertex $v\in S$ has at least as many neighbours (including itself) in $S$ as it has 
in $V\setminus S$. 
We consider the notion of global minimality in this paper. We are interested in globally
 minimal defensive alliance of maximum size. This 
problem is known to be NP-hard  but its parameterized complexity remains open 
until now.  We enhance our understanding of the problem from the 
viewpoint of parameterized complexity by showing that 
the {\sc Globally Minimal Defensive Alliance}  problem is FPT parameterized by 
the neighbourhood diversity of the input graph. The result for neighborhood diversity implies that the problem is FPT parameterized by  vertex cover number also. 
We prove that the  problem parameterized by the vertex cover number of the input graph does not admit a polynomial compression unless coNP $\subseteq$ NP/poly.
We show that the problem is W[1]-hard parameterized by a wide range of fairly restrictive structural parameters such as the feedback vertex set number, pathwidth, treewidth and treedepth. 
We also proved that, given a vertex $r \in V(G)$, deciding if $G$ has a globally minimal defensive alliance of any size containing vertex $r$ is NP-complete.

\end{abstract}


\section{Introduction}
During the last 20 years, the {\sc Defensive Alliance} problem has been studied
extensively. A defensive alliance in an undirected graph is a non-empty set of vertices 
with the property that each vertex  has at least as many neighbours  in the alliance
(including itself) as outside the alliance.  In 2000, Kristiansen, Hedetniemi, and Hedetniemi \cite{kris}
introduced defensive, offensive, and powerful alliances, and further studied by Shafique \cite{HassanShafique2004PartitioningAG}
and other authors \cite{BAZGAN2019111,BLIEM2018334,small,Cami2006OnTC,Enciso2009AlliancesIG,Fernau,FERNAU2009177,ICDCIT2021,Lindsay,ROD,SIGARRETA20091687,SIGARRETA20061345,SIGA}. In this paper, we will focus on defensive alliances. 
The theory of alliances in graphs have been studied 
intensively \cite{Cami2006OnTC,10.5614/ejgta.2014.2.1.7,frick} both from a combinatorial and from a computational perspective. 
As mentioned in \cite{BAZGAN2019111}, the 
focus has been mostly on finding small alliances, although studying large
alliances do not only make a lot of sense from the original motivation of these notions, 
but was actually also delineated in the very first papers on alliances \cite{kris}.

Note that a defensive alliance is not a hereditary property, that is, a subset of 
defensive alliance is not necessarily a  defensive alliance.  
Shafique \cite{HassanShafique2004PartitioningAG} called 
an alliance a {\it locally minimal alliance} if the set obtained by removing any vertex of the
alliance is not an alliance. Bazgan  et al. \cite{BAZGAN2019111} considered another notion 
of alliance that they called a {\it globally minimal alliance} which has the property that 
no proper subset is an alliance. In this paper we are interested in globally minimal 
alliances of maximum size. Clearly, the motivation is that big communities where every member
still matters somehow are of more interest than really small communities. Also, there is a 
general mathematical interest in such type of problems, see \cite{Manlove1998MinimaximalAM}.

Throughout this article, $G=(V,E)$ denotes a finite, simple and undirected graph of order $|V|=n$. 
For a non-empty subset $S\subseteq V$ and a vertex $v\in V(G)$, 
let $N_S(v)=\{ u\in S~:~ (u,v)\in E(G)\}$, $N_S[v]=N_S(v)\cup \{v\}$,  and $d_S(v)$ denote
its open neighborhood, closed neighborhood, and degree respectively in $S$.
The complement of the vertex set $S$ in $V$ is denoted by $S^c$.
\begin{definition}\rm
A non-empty set $S\subseteq V$ is a defensive alliance in $G$ if for each $v\in S$, 
$|N[v]\cap S|\geq |N(v)\setminus S|$, or equivalently,   $d_S(v)+1\geq d_{S^c}(v)$. 
\end{definition}
\noindent A vertex $v\in S$ is said to be {\it protected} if $d_S(v)+1\geq d_{S^c}(v)$. 
A non-empty set $S\subseteq V$ is a defensive alliance if every vertex
in $S$ is protected. 
\begin{definition}\rm
A vertex $v\in S$ is said to be {\it marginally protected} 
if it becomes unprotected when any of its neighbour in $S$ is moved from $S$ to $V\setminus S$.
A vertex $v\in S$ is said to be {\it strongly protected} if it remains protected
when any of its neighbours is moved from $S$ to $V\setminus S$. 
\end{definition}

\begin{definition}\rm 
A defensive alliance $S$ is called a {\it locally minimal defensive alliance}
if for any $v\in S$, $S\setminus \{v\}$ is not a defensive alliance.
\end{definition}

\begin{definition}\rm 
A defensive alliance $S$ is called a {\it globally minimal defensive alliance} 
if no proper 
subset is a defensive alliance.
\end{definition}

\noindent In literature, a defensive alliance $S$ is called {\it global defensive alliance} if $S$ is a dominating set. 
It is to be noted  that globally minimal defensive alliance is different from global defensive alliance. 

\begin{observation}\label{onedegree}\rm
 Let $S$ be a globally minimal defensive alliance of size at least two in $G$.
 Then $S$ can never contain a vertex of degree one. 
 \end{observation}
 \noindent This can be proved by contradiction. Suppose $S$ contains a vertex $v$ of degree one. 
 Note that $\{v\}$ is a proper subset of $S$ and it is a defensive alliance, a contradiction to the fact that
 $S$ is a globally minimal defensive alliance. \\
 
\noindent A  defensive alliance $S$ is connected if the subgraph $G[S]$ induced by 
$S$ is connected. 
Notice that any globally minimal defensive alliance is always connected. 

\begin{observation}\label{OBS2}\rm
If a non-empty set $S \subseteq V$ is connected and  each $v\in S$ is marginally protected, then $S$ is a globally minimal defensive alliance. 
 \end{observation} 

\noindent Note that although the conditions of this observation are sufficient to 
assure that $S$ is a globally minimal defensive alliance, they  are certainly  not necessary. For example, consider the graph $G$ shown in Figure \ref{OBS2}. Note that $S=\{x,y,u_1,u_2\}$ is
a globally minimal defensive alliance, but $u_1$ and $u_2$ are not marginally protected.

\begin{figure}[ht]
     \centering
 \begin{tikzpicture}[scale=0.8]
\vertex [fill=pink] (x) at (0,-0.5) [label=below:${x}$] {};
\vertex [fill=pink] (y) at (2,-0.5) [label=below:$y$] {};
\vertex [fill=pink] (u1) at (-1,1) [label=above:$u_{1}$] {};
\vertex [fill=pink] (u2) at (0,1) [label=above:$u_{2}$] {};
\vertex [fill=white](u3) at (1,1) [label=above:$u_{3}$] {};
\vertex [fill=white](u4) at (2,1) [label=above:$u_{4}$] {};
\vertex [fill=white](u5) at (3,1) [label=above:$u_{5}$] {};

\draw(x)--(y);
\draw(x)--(u1);
\draw(x)--(u2);
\draw(x)--(u3);
\draw(x)--(u4);
\draw(x)--(u5);
\draw(y)--(u1);
\draw(y)--(u2);
\draw(y)--(u3);
\draw(y)--(u4);
\draw(y)--(u5);
 \end{tikzpicture}
     \caption{$S=\{x,y,u_1,u_2\}$ is a globally minimal defensive alliance in $G$.}
     \label{counterexample}
 \end{figure}
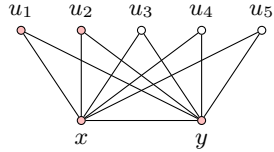

\noindent 
In this paper, we consider {\sc Globally Minimal Defensive Alliance}
problem under structural parameters. We define the problem as follows:
    \vspace{3mm}
    \\
    \fbox
    {\begin{minipage}{33.7em}\label{FFVS }
       {\sc Globally Minimal Defensive Alliance}\\
        \noindent{\bf Input:} An undirected graph $G=(V,E)$ and an  integer $k \geq 2$.
    
        \noindent{\bf Question:} Is there a globally minimal defensive alliance $S\subseteq V(G)$ such that 
        $|S|\geq  k $?
    \end{minipage} }
  \vspace{3mm}

\noindent For the standard concepts in parameterized complexity, see the recent textbook by Cygan et al. \cite{marekcygan}.    \noindent  For standard notations and definitions in graph theory, we refer to West \cite{west}. The graph
parameters we explicitly use in this paper are  feedback vertex set number, pathwidth, treewidth and treedepth.  

    \begin{definition} {\rm
        For a graph $G = (V,E)$, the parameter {\it feedback vertex set} is the cardinality of the smallest set $S \subseteq V(G)$ such that the graph $G-S$ is a forest and it is denoted by $fvs(G)$.}
    \end{definition}

\noindent We now review the concept of a tree decomposition, introduced by Robertson and Seymour in \cite{Neil}.
Treewidth is a  measure of how “tree-like” the graph is.
\begin{definition}\rm \cite{Downey} A {\it tree decomposition} of a graph $G=(V,E)$  is a tree $T$ together with a 
collection of subsets $X_t$ (called bags) of $V$ labeled by the vertices $t$ of $T$ such that 
$\bigcup_{t\in T}X_t=V $ and (1) and (2) below hold:
\begin{enumerate}
			\item For every edge $(u,v) \in E(G)$, there  is some $t$ such that $\{u,v\}\subseteq X_t$.
			\item  (Interpolation Property) If $t$ is a vertex on the unique path in $T$ from $t_1$ to $t_2$, then 
			$X_{t_1}\cap X_{t_2}\subseteq X_t$.
		\end{enumerate}
	\end{definition}
	
	
\begin{definition}\rm \cite{Downey} The {\it width} of a tree decomposition is
the maximum value of $|X_t|-1 $ taken over all the vertices $t$ of the tree $T$ of the decomposition.
The treewidth $tw(G)$ of a graph $G$  is the  minimum width among all possible tree decomposition of $G$.
\end{definition} 

\begin{definition}\rm 
    If the tree $T$ of a tree decomposition is a path, then we say that the tree decomposition 
    is a {\it path decomposition}, and use {\it  pathwidth} in place of treewidth. 
\end{definition}

A rooted forest is a disjoint union of rooted trees. Given a rooted forest $F$, its \emph{transitive closure} is a graph $H$ in which $V(H)$ contains all the nodes of the rooted forest, and $E(H)$ contain an edge between two vertices only if those two vertices form an ancestor-descendant pair in the forest $F$.

   \begin{definition}
        {\rm  The {\it treedepth} of a graph $G$ is the minimum height of a rooted forest $F$ whose transitive closure contains the graph $G$. It is denoted by $td(G)$.}
    \end{definition}

\subsection{Our Results}
The goal of this paper is to provide new insight into the complexity of 
{\sc Globally Minimal Defensive Alliance} parameterized by the structure of the input graph. In this paper, we prove the following results:
 \begin{itemize}
     \item The {\sc Globally Minimal  Defensive Alliance} problem is fixed-parameter tractable when parameterized by  the neighbourhood diversity.
     \item The   problem parameterized by the vertex cover number of the input graph does not admit a polynomial 
compression unless coNP $\subseteq$ NP/poly.
\item The problem is W[1]-hard parameterized by a wide range of fairly restrictive structural parameters such as the feedback vertex set number, pathwidth, treewidth and treedepth.
\item Given a vertex $r \in V(G)$, deciding if $G$ has  a globally minimal defensive alliance containing vertex $r$ is NP-complete.
 \end{itemize}

\subsection{Known Results} The decision version for several types of alliances have been shown to be NP-complete. 
For an integer $r$, a nonempty set $S\subseteq V(G)$ is a {\it defensive $r$-alliance} if for each 
$v\in S$, $|N(v)\cap S|\geq |N(v)\setminus S|+r$. A set is a defensive alliance if it is a defensive 
$(-1)$-alliance. A defensive $r$-alliance $S$ is {\it global} if $S$ is a dominating set. 
 The defensive $r$-alliance problem   is NP-complete for any $r$ \cite{SIGARRETA20091687}. The defensive alliance problem is 
 NP-complete even when restricted to split, chordal and bipartite graph \cite{Lindsay}. 
 For an integer $r$, a nonempty set $S\subseteq V(G)$ is an {\it offensive $r$-alliance} if for each 
$v\in N(S)$, $|N(v)\cap S|\geq |N(v)\setminus S|+r$. An offensive 1-alliance is called an offensive
alliance.  An offensive $r$-alliance $S$ is {\it global} if $S$ is a dominating set. 
 Fernau et al. showed that the offensive $r$-alliance and global 
 offensive $r$-alliance problems are NP-complete for any fixed $r$ \cite{FERNAU2009177}. 
 They also proved that for $r>1$, $r$-offensive alliance is NP-hard, even when restricted to 
 $r$-regular planar graphs.  There are polynomial time algorithms for finding minimum alliances
 in trees \cite{CHANG2012479,Lindsay}. 
Bliem and Woltran \cite{BLIEM2018334} proved that deciding if a graph contains a defensive alliance of size at most
$k$ is W[1]-hard when parameterized by treewidth of the input graph. Bazgan et al. \cite{BAZGAN2019111} proved that deciding if a graph contains a globally minimal strong defensive alliance of size at least $k$ is NP-complete, even for cubic graphs. Moreover, deciding if a graph contains a globally minimal defensive alliance of size at least $k$ is NP-complete, even for graphs of degree 3 or 4 \cite{BAZGAN2019111}.
 
\section{FPT algorithm parameterized by neighbourhood diversity}\label{ndsection}

In this section, we present an FPT algorithm for {\sc Globally Minimal Defensive Alliance}  problem parameterized 
 by neighbourhood diversity. 
  We say two vertices $u$ and $v$ have the same type if and only if 
 $N(u)\setminus \{v\}=N(v)\setminus \{u\}$. The relation of having the same type 
 is an equivalence  relation. The idea of neighbourhood diversity is based on this 
 type structure. 
  \begin{definition} \rm \cite{Lampis} 
        The neighbourhood diversity of a graph $G=(V,E)$, denoted by ${\tt nd}(G)$, is the least integer $k$ for which we can partition the set $V$ of vertices  into $k$ classes, such that all vertices in each class have the 
        same type.
    \end{definition}
    
    If neighbourhood diversity of a graph is bounded by an integer $k$, then there exists 
    a partition $\{ C_1, C_2,\ldots, C_k\}$ of $V(G)$ into $k$ type  classes. It is known that 
    such a minimum partition can be found in linear time using fast modular decomposition algorithms \cite{Tedder}. 
    Notice
    that each type class  could either be a clique or an independent set by definition.  
     For algorithmic 
    purpose it is often useful to consider a {\it type graph} $H$ of  graph $G$, where
    each vertex of $H$ is a type class in $G$, and two vertices $C_i$ and $C_j$ are adjacent iff
    there is complete bipartite clique between these type classes in $G$. It is not difficult to see that
    there will be either a complete bipartite clique or no edges between any two type classes. 
      The key property of graphs of
  bounded neighbourhood diversity is that their type graphs have bounded size.  
  In this section, we prove the following theorem:
    \begin{theorem}\label{theoremnd1}
       The {\sc Globally Minimal  Defensive Alliance} problem is fixed-parameter tractable when parameterized by  the neighbourhood diversity.
    \end{theorem}
    
Let $G$ be a connected graph such that ${\tt nd}(G)=k$. Let $C_1,\ldots,C_k$ be the partition 
    of $V(G)$ into sets of type classes. We assume $k\geq 2$ since otherwise the problem 
     becomes trivial. Let $x_i=|C_i\cap S|$ where $S$ is a globally minimal defensive alliance.
     We define $I_0=\{ C_i~|~ x_i=0\}$, $I_1=\{ C_i~|~x_i=1\}$ and 
    $I_{2}=\{ C_i~|~x_i\geq 2 \}$. We next guess if $C_i$ belongs to $ I_0$, $I_1$, or $ I_{2}$.
    There are at most $3^k$ possibilities as each $C_i$ has three options: either in 
    $I_0$, $I_1$, or $I_{2}$. We reduce the problem of finding a globally minimal defensive alliance
    to an integer linear programming  optimization with $k$ variables. 
    Since integer linear programming is fixed parameter tractable when parameterized by 
    the number of variables \cite{lenstra}, we conclude that our problem is FPT when parameterized by 
    the neighbourhood diversity $k$. \\

     \noindent  {\bf ILP formulation:} Our goal here is to find a largest globally
        minimal defensive alliance $S$ of  $G$, with $C_i\cap S= \emptyset$ when $C_i\in I_0$,
        $|C_i\cap S|=1$ when $C_i \in  {I_1}$, and $|C_i\cap S|\geq 2$ when $C_i \in  {I_{2}}$
        where $I_0,I_1, I_{2}$ are given. 
        For each $C_i$, we associate a variable $x_i$ that indicates
        $|S\cap C_i|=x_i$. As the vertices in $C_i$ have the same neighbourhood, 
        the variables $x_i$ determine
        $S$ uniquely, up to isomorphism.  The objective  here is to 
        maximize $\sum\limits_{C_i\in I_1\cup I_2}{x_i}$ under the conditions given below. 
        Let $\mathcal{C}$ be a subset of $I_1\cup I_2$ consisting of all type classes which are cliques; and
        $\mathcal{I}={I_1\cup I_2}\setminus \mathcal{C}$.  
        We consider two cases:\\
        
        \noindent{\bf  Case 1:} Suppose $v \in C_j$ where $C_j\in \mathcal{I}$.
         Then the number of neighbours of $v$ in $S$, including itself, is $1+ d_S(v) =\sum\limits_{C_i\in N_H(C_j) \cap {(I_1\cup I_2)}}{x_i}$.
 Note that if $C_i\in N_H(C_j)  \cap (I_1\cup I_2)$, then only $x_i$ 
        vertices of $C_i$ are in $S$ and the the remaining $n_i-x_i$ vertices of 
        $C_i$ are outside $S$. 
        The number of neighbours of $v$ outside $S$ is  
    $\sum\limits_{C_i\in N_H(C_j) \cap {(I_1\cup I_2)}}{(n_i-x_i)} + \sum\limits_{ C_i\in N_H(C_j)\cap I_0}{n_i}$.
  Therefore,  a vertex $v$ from an independent type class $C_j \in \mathcal{I}$ 
is protected if and only if 
\begin{align*}
  1+\sum\limits_{C_i\in N_H(C_j) \cap {(I_1\cup I_2)}}{x_i}\geq \sum\limits_{C_i\in N_H(C_j) \cap {(I_1\cup I_2)}}{(n_i-x_i)} + \sum\limits_{ C_i\in N_H(C_j)\cap {I_0}}{n_i},  \end{align*}
  or equivalently, 
  \begin{align}
  1+ \sum\limits_{C_i\in N_H(C_j) \cap (I_1\cup I_2)}{2x_i} \geq \sum\limits_{C_i\in N_H(C_j)}{n_i}
  \end{align}

        \noindent{\bf Case 2:} Suppose $v\in C_j$ where  $C_j\in \mathcal{C}$.   The number of neighbours of $v$ in $S$, including itself, is     
          $ \sum\limits_{C_i\in N_H[C_j]\cap (I_1\cup I_2)}{x_i}$.  
         The number of neighbours of $v$ outside $S$ is  
           $\sum\limits_{C_i\in N_H[C_j]\cap (I_1\cup I_2)}{(n_i-x_i)} +\sum\limits_{C_i\in N_H[C_j]\cap {I_0}}{n_i}$. 
         Thus a vertex  $v$ from a clique type class $C_j \in \mathcal{C}$ is protected if and only if  
         $d_S(v)+1\geq d_{S^c}(v) $, that is, 
         \begin{align*}
           \sum\limits_{C_i\in N_H[C_j]\cap (I_1\cup I_2) }{x_i} \geq \sum\limits_{C_i\in N_H[C_j]\cap (I_1\cup I_2)}{(n_i-x_i)} +\sum\limits_{C_i\in N_H[C_j]\cap I_0}{n_i},  
         \end{align*}
         or equivalently, 
         \begin{align}
           \sum\limits_{C_i\in N_H[C_j]\cap (I_1\cup I_2)}{2x_i}\geq \sum\limits_{C_i\in N_H[C_j]}{n_i}. 
         \end{align}
         \\
        
        Let ${\bf x}=(x_1,\ldots,x_k)$ be the vector that correspond to the set $S\subseteq V(G)$. 
We want to make sure that the vector ${\bf x}=(x_1,\ldots,x_k)$ or the set $S$ forms a defensive alliance,
     but no proper subset of $S$ forms a defensive alliance. We now characterize all proper subsets 
     of $S$ in terms of $k$-length vectors. 
      We define a new variable $y_i$ as follows: $0<y_i< x_i$ for all $i$.
  Let $L({\bf x})$ be the set of all length $k$ vectors where the $i$th entry be either
  $0$, $y_i$ or $x_i$. 
  Note that  each vector in $L({\bf x})$ 
   represents a proper subset of  $S$ unless the $i$th entry is $x_i$ for all $i$.  
   The number of vectors in $L({\bf x})$ is $\prod\limits_{i=1}^k{(x_i+1)}$. 
  We define another set $L^{\prime}({\bf x})$ as follows: let $L^{\prime}({\bf x})$ be the set of 
  all length $k$ vectors where the $i$th entry is either
  $0$, $x_i-1$ or $x_i$; note that $x_i-1$ is  possible only if $x_i\geq 2$, that is, $C_i\in I_2$. 
  Clearly, $L^{\prime}({\bf x}) \subseteq L({\bf x})$ and $L^{\prime}({\bf x})$ has at most
  $3^k$ vectors. To prove Theorem \ref{theoremnd1} we need the following lemma. 
  

  \begin{lemma}\rm 
  Let ${\bf x}=(x_1,\ldots,x_k)$ be the vector that represent $S\subseteq V(G)$. If no vector in $L^{\prime}({\bf x})$ forms a defensive alliance then no vector in
  $L({\bf x})$ forms a defensive alliance.
  \end{lemma}
  \proof  Assume, for the sake of contradiction, that  ${\bf x}_1\in L({\bf x})$ forms a defensive 
  alliance. Without loss of generality, let ${\bf x}_1=(y_1,x_2-1, y_3,\ldots, x_k)$, then we obtain the vector
  ${\bf x}^{\prime}_1=(x_1-1,x_2-1, x_3-1,\ldots, x_k) \in L^{\prime}({\bf x})$ from ${\bf x}_1$ by replacing $y_i$ by
  $x_i-1$ for all $i$. 
  As ${\bf x}^{\prime}_1 \in L^{\prime}({\bf x})$, we know ${\bf x}^{\prime}_1$ does not form a defensive alliance. 
  This means, there is a vertex $u\in C_i$ which is not protected in ${\bf x}^{\prime}_1$ 
  (assume that the $i$th entry of 
  ${\bf x}^{\prime}_1$ is non-zero). We observe that the number of neighbours of 
  $u$ in ${\bf x}_1$, is less than or equal to  the number 
  of neighbours in ${\bf x}^{\prime}_1$. In other words, $u$ is not protected  
  in ${\bf x}_1$ either, a contradiction to the assumption that ${\bf x}_1\in L(\bf x)$ forms a defensive 
  alliance. This proves the lemma. \qed

\noindent In order to  ensure that $S$ is a globally minimal defensive alliance, 
we check ${\bf x}=(x_1,x_1,\ldots,x_k)$ forms a 
  defensive alliance but none of the vectors in $L^{\prime}({\bf x})$ forms a defensive
  alliance. Let ${\bf x}^{\prime}_1, {\bf x}^{\prime}_2, \ldots,{\bf x}^{\prime}_{3^k} $ be  
  the vectors in $L^{\prime}({\bf x})$. 
  We make guesses in two phases. In the first phase, we guess if $C_i$ belongs to $I_0,I_1$ or $I_2$.
  There are at most $3^k$ possibilities as each $C_i$ has three options: either in $I_0,I_1$ or $I_2$.
  In the second phase, we guess if an unprotected vertex of ${\bf x}^{\prime}_i$ belongs to 
  type class either $C_1, \ldots, C_{k-1}$ or $C_k$.  We define 
  $$ R_j=\Big\{ {\bf x}^{\prime}_i \in L^{\prime}({\bf x}) ~|~ \text{ an unprotected vertex
  of } {\bf x}^{\prime}_i \text{ is in type class } C_j\Big\}.$$
  There are at most
  $k^{3^k}$ possibilities as each ${\bf x}^{\prime}_i$ has at most $k$ options: 
  $R_1, R_2, \ldots, R_{k}$.
   If $C_j$ is an independent type class, then it contains an unprotected vertex if, 
  $$1+ \sum\limits_{C_i\in N_H(C_j) \cap (I_1\cup I_2)}{2x_i^{\prime}} < \sum\limits_{C_i\in N_H(C_j)}{n_i}.$$  
  If $C_j$ is a clique type class, then it contains an unprotected vertex if,
 $$\sum\limits_{C_i\in N_H[C_j]\cap (I_1\cup I_2)}{2x_i^{\prime}}< \sum\limits_{C_i\in N_H[C_j]}{n_i}.$$  
 We now formulate ILP formulation of globally minimal defensive alliance problem, for 
 given $I_0,I_1,I_2$ and $R_1,R_2,\ldots,R_k$. There are at most $3^kk^{3^k}$ ILPs:\\
 
\noindent       
\vspace{3mm}
    \fbox
    {\begin{minipage}{33.7em}\label{Min-FFVS}      
\begin{equation*}
\begin{split}
&\text{Maximize } \sum\limits_{C_i\in I_1\cup I_2}{x_i}\\
&\text{Subject to~~~} \\
&                                         x_{i}=1 \text{ for all }i ~:~ C_i\in  I_1;\\
&                                         x_{i} \in \{2,\ldots, |C_i| \} \text{ for all }i ~:~ C_i\in  I_2\\
&1+ \sum\limits_{C_i\in N_H(C_j) \cap (I_1\cup I_2)}{2x_i} \geq \sum\limits_{C_i\in N_H(C_j)}{n_i}, ~~\text{for all } C_j\in \mathcal{I}, \\
&\sum\limits_{C_i\in N_H[C_j]\cap (I_1\cup I_2)}{2x_i}\geq \sum\limits_{C_i\in N_H[C_j]}{n_i},~~\text{for all } C_j\in \mathcal{C},  \\
& \text{ for } j=1 \text{ to } k;\\
&1+ \sum\limits_{C_i\in N_H(C_j) \cap (I_1\cup I_2)}{2x_i^{\prime}} < \sum\limits_{C_i\in N_H(C_j)}{n_i}, \forall ~{\bf x}^{\prime}_i\in R_j; C_j \text{ is an independent class} \\
&\sum\limits_{C_i\in N_H[C_j]\cap (I_1\cup I_2)}{2x_i^{\prime}}< \sum\limits_{C_i\in N_H[C_j]}{n_i},\forall~ {\bf x}^{\prime}_i\in R_j; C_j \text{ is  a clique class} \\
\end{split}
\end{equation*}  
  \end{minipage} }

\noindent {\bf Solving the ILP:}
Lenstra \cite{lenstra} showed that the feasibility version of {\sc $k$-ILP} is FPT with 
running time doubly exponential in $k$, where $k$ is the number of variables. 
Later, Kannan \cite{kannan} proved an algorithm for {\sc $k$-ILP} running in time $k^{O(k)}$.
In our algorithm, we need the optimization version of {\sc $k$-ILP} rather than 
the feasibility version. We state the minimization version of {\sc $k$-ILP}
as presented by Fellows et al. \cite{fellows}. \\

\noindent {\sc $k$-Variable Integer Linear Programming Optimization ($k$-Opt-ILP)}: Let matrices $A\in \ Z^{m\times k}$, $b\in \ Z^{k\times 1}$ and 
$c\in \ Z^{1\times k}$ be given. We want to find a vector $ x\in \ Z ^{k\times 1}$ that minimizes the objective function $c\cdot x$ and satisfies the $m$ 
inequalities, that is, $A\cdot x\geq b$.  The number of variables $k$ is the parameter. 
Then they showed the following:

\begin{lemma}\rm \label{ilp}\cite{fellows}
{\sc $k$-Opt-ILP} can be solved using $O(k^{2.5k+o(k)}\cdot L \cdot log(MN))$ arithmetic operations and space polynomial in $L$. 
Here $L$ is the number of bits in the input, $N$ is the maximum absolute value any variable can take, and $M$ is an upper bound on 
the absolute value of the minimum taken by the objective function.
\end{lemma}

In the formulation for {\sc Globally Minimal Defensive Alliance} problem, we have at most $k$ variables. The value of objective function is bounded by $n$ and the value of any variable 
in the integer linear programming is also bounded by $n$. The constraints can be represented using 
$O(k^2 \log{n})$ bits. Lemma \ref{ilp} implies that we can solve the problem with one guess in FPT time. 
There are at most $3^kk^{3^k}$ guesses, and the ILP formula for each guess can be solved in FPT time. Thus 
Theorem \ref{theoremnd1} holds.

 \section{No polynomial kernel parameterized by vertex cover number}
 A set $C\subseteq V$ is a vertex cover of $G=(V,E)$ if each edge $e\in E$ has at least one endpoint 
 in $X$. The minimum size of a vertex cover in $G$ is the {\it vertex cover number} of $G$, denoted by $vc(G)$. The problem is FPT parameterized by neighbourhood diversity implies that it is FPT parameterized by vertex cover number $vc$.
 In this section we prove the following 
 kernelization hardness of  the {\sc Globally Minimal Defensive Alliance} problem.
 
  \begin{theorem}\label{pptGMDA}\rm
The {\sc Globally Minimal Defensive Alliance}  problem parameterized by the vertex cover number of the input graph does not admit a polynomial 
compression unless coNP $\subseteq$ NP/poly.
 \end{theorem}
 
 \noindent To prove Theorem \ref{pptGMDA}, we give a polynomial parameter transformation (PPT) from the well-known 
 {\sc Red Blue Dominating Set} problem (RBDS) to {\sc Globally Minimal Defensive Alliance} parameterized by vertex cover number. 
 Recall that in RBDS we are given a bipartite graph $G=(T\cup S,E)$ and an integer $k$, and we are 
 asked whether there exists a vertex set $X\subseteq S$ of size at most $k$ such that every vertex in $T$
 has at least one neighbour in $X$. We also refer to the vertices of $T$ as {\it terminals} and 
 to the vertices of $S$ as {\it sources} or {\it nonterminals}. The following theorem is known:
 
 \begin{theorem}\label{RBDS} \rm \cite{fomin_lokshtanov_saurabh_zehavi_2019}
 RBDS parameterized by $|T|$ does not admit a polynomial compression unless coNP $\subseteq$ NP/poly.
 \end{theorem}
 
\subsection{Proof of Theorem \ref{pptGMDA}} 
\begin{figure}
    \centering
    \begin{tikzpicture}[scale=0.7]
\draw[yellow,thick] (0,0) ellipse (0.4cm and 1.7cm);

\draw[yellow,thick] (3,0) ellipse (0.4cm and 2.5cm);

\node[circle,draw, inner sep=0 pt, minimum size=0.1cm](u) at (0,1.3) [label=below:$u$]{};

\node[circle,draw, inner sep=0 pt, minimum size=0.1cm](u1) at (0,2.3) [label=right:$u_{1}$]{};

\node[circle,red,draw, inner sep=0 pt, minimum size=0.1cm](l1s) at (4,3.3) [label=below:]{};

\node[circle,red,draw, inner sep=0 pt, minimum size=0.1cm](l2s) at (4.5,3.3) [label=below:]{};

\node[circle,red,draw, inner sep=0 pt, minimum size=0.1cm](l3s) at (5,3.3) [label=below:]{};

\node[red](t1square) at (4.5,3.4) [label=above:$T_{1}^{\square}$]{};

\draw[red, thick] (3.8,3) rectangle (5.2,3.6);


\node[circle,draw,color=red, inner sep=0 pt, minimum size=0.1cm](t2s) at (2,3.3) [label=below:]{$T_2^{\square}$};

\node[circle,draw,color=red, inner sep=0 pt, minimum size=0.1cm](vu1) at (-1,2.8) [label=below:]{$V_{u_{1}}^{\square}$};

\node[thick, circle,draw,color=blue, inner sep=0 pt, minimum size=0.1cm](vu0) at (0,3.3) [label=below:]{$V_{u}^{\triangle}$};

\node[circle,draw, inner sep=0 pt, minimum size=0.1cm](u2) at (0,4.3) [label=right:$u_{2}$]{};

\node[circle,draw,color=red, inner sep=0 pt, minimum size=0.1cm](vu2) at (-1,4.8)
[label=below:]{$V_{u_{2}}^{\square}$};

\node[circle,draw, inner sep=0 pt, minimum size=0.1cm](v) at (0,-1.3) [label=above:$v$]{};

\node[circle,draw, inner sep=0 pt, minimum size=0.1cm](v1) at (0,-2.3) [label=right:$v_{1}$]{};

\node[circle,draw,color=red, inner sep=0 pt, minimum size=0.1cm](vv1) at (-1,-2.8) [label=below:]{$V_{v_{1}}^{\square}$};

\node[thick, circle,draw,color=blue, inner sep=0 pt, minimum size=0.1cm](vv0) at (0,-3.3) [label=below:]{$V_{v}^{\triangle}$};

\node[circle,draw, inner sep=0 pt, minimum size=0.1cm](v2) at (0,-4.3) [label=right:$v_{2}$]{};

\node[circle,draw,color=red, inner sep=0 pt, minimum size=0.1cm](vv2) at (-1,-4.8) [label=below:]{$V_{v_{2}}^{\square}$};

\node[circle,draw, inner sep=0 pt, minimum size=0.1cm](b) at (-3,-1.3) [label=above:$b$]{};

\node[circle,draw,color=red, inner sep=0 pt, minimum size=0.1cm](vb) at (-4,-1.3) [label=below:]{$V_{b}^{\square}$};

\node[circle,draw, inner sep=0 pt, minimum size=0.1cm](a) at (-3,1.3) [label=above:$a$]{};

\node[circle,draw,color=red, inner sep=0 pt, minimum size=0.1cm](va) at (-4,1.3) [label=below:]{$V_{a}^{\square}$};

\node[thick, circle,draw, inner sep=0 pt, minimum size=0.1cm](vua) at (-1.5,1.8) [label=below:]{$V_{u}^{a}$};

\node[thick, circle,draw,  inner sep=0 pt, minimum size=0.1cm](vva) at (-2.5,0.3) [label=below:]{$V_{v}^{a}$};

\node[circle,draw, inner sep=0 pt, minimum size=0.1cm](z1) at (3,2.1) [label=below:]{};
\node[circle,draw, inner sep=0 pt, minimum size=0.1cm](z2) at (3,0.9) [label=below:]{};
\node[circle,draw, inner sep=0 pt, minimum size=0.1cm](z3) at (3,-0.9) [label=below:]{};
\node[circle,draw, inner sep=0 pt, minimum size=0.1cm](z4) at (3,-2.1) [label=below:]{};

\node[circle,draw, inner sep=0 pt, minimum size=0.1cm](x) at (4.5,0) [label=above:$x$]{};

\node[thick, circle,draw,color=blue, inner sep=0 pt, minimum size=0.1cm](vx0) at (5.5,0) [label=below:]{$V_{x}^{\triangle}$};

\path 
(vu0) edge[pink,thick] (t2s)
(vv0) edge[pink,thick] (t2s)
(vx0) edge[pink,thick] (t2s)
(t2s) edge[pink,thick] (z1)
(t2s) edge[pink,thick] (z2)
(t2s) edge[pink,thick] (z3)
(t2s) edge[pink,thick] (z4);

\node[circle,draw, inner sep=0 pt, minimum size=0.1cm](x1) at (6.5,0) [label=above:$x'$]{};

\node[circle,draw,color=red, inner sep=0 pt, minimum size=0.1cm](vx1) at (4.5,-1) [label=below:]{$V_{x}^{\square}$};

\node[circle,draw,color=red, inner sep=0 pt, minimum size=0.1cm](vx2) at (7.5,0) [label=below:]{$V_{x'}^{\square}$};

\node(T) at (0.1,-1.7) [label=right:$T$]{};
\node(S) at (3,-3.3) [label=above:$S$]{};

\path

(z1)edge[cadetgrey](l1s)
(z1)edge[cadetgrey](l2s)
(z1)edge[cadetgrey](l3s)
(z2)edge[cadetgrey](l1s)
(z2)edge[cadetgrey](l2s)
(z2)edge[cadetgrey](l3s)
(z3)edge[cadetgrey](l1s)
(z3)edge[cadetgrey](l2s)
(z4)edge[cadetgrey](l1s)
(z4)edge[cadetgrey](l2s);

\draw(u)--(u1);
\draw(u)--(vua);
\draw(u)--(a);
\draw(u)--(b);
\draw(u)--(z1);
\draw(u)--(z2);
\draw(u)--(z3);
\draw(a)--(va);
\draw(b)--(vb);
\draw(a)--(vva);
\draw(a)--(vua);
\draw(v)--(v1);
\draw(v)--(vva);
\draw(v)--(a);
\draw(v)--(b);
\draw(v)--(z1);
\draw(v)--(z2);
\draw(v)--(z4);
\draw(u1)--(vu0);
\draw(u2)--(vu0);
\draw(u1)--(vu1);
\draw(u2)--(vu2);
\draw(v1)--(vv0);
\draw(v2)--(vv0);
\draw(v1)--(vv1);
\draw(v2)--(vv2);
\draw(x)--(z1);
\draw(x)--(z2);
\draw(x)--(z3);
\draw(x)--(z4);
\draw(x)--(vx1);
\draw(x)--(vx0);
\draw(x1)--(vx2);
\draw(x1)--(vx0);

\draw (b) .. controls(0,-2.7) .. (3,-3.3)..controls(4,-3.5).. (x1);

    \end{tikzpicture}
    \caption{PPT from RBDS to {\sc Globally Minimal Defensive Alliance}}
    \label{fig:RBDS}
\end{figure}
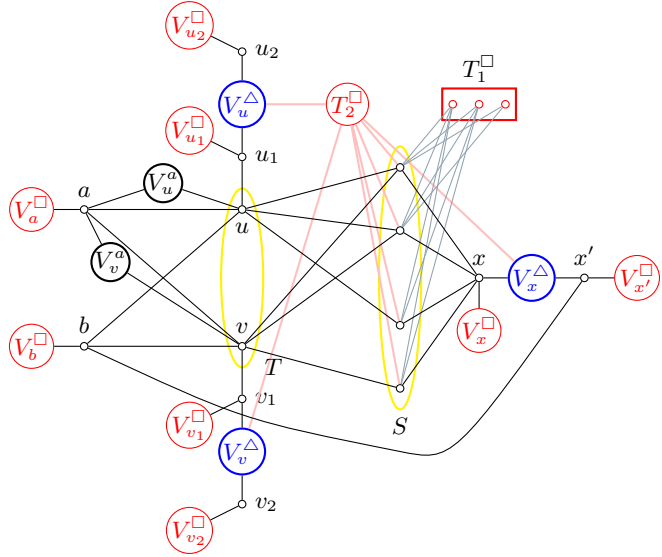

By Theorem \ref{RBDS}, RBDS parameterized by $|T|$ does not admit a polynomial compression unless coNP $\subseteq$ NP/poly.
To prove Theorem \ref{pptGMDA}, we give a PPT from RBDS parameterized by $|T|$ to {\sc Globally Minimal Defensive Alliance} parameterized by 
the vertex cover number.
 Given an instance $I=(G = (T\cup S, E),k)$ of RBDS, we construct  an instance 
$I'=(G',k')$ of {\sc Globally Minimal Defensive Alliance} as follows.  See
Fig. \ref{fig:RBDS} for an illustration. 

\begin{itemize}
\item We introduce two new vertices $x$ and $x'$, a set $V_{x}^{\triangle}$ of $4n$ vertices, a set $V_{x}^{\square}$ of $|S|-k+4n+1$ vertices, and a set $V_{x'}^{\square}$ of $4n+1$ vertices. Make $x$  adjacent to every vertex of $V_{x}^{\triangle} \cup S \cup V_{x}^{\square}$ and  make $x'$  adjacent to every vertex of $V_{x}^{\triangle} \cup V_{x'}^{\square}$.
  \item For every vertex $u\in T$, we introduce two vertices $u_{1}$ and $u_{2}$,
  a set $V_{u}^{\triangle}$ of $4n$ vertices, a set $V_{u_{1}}^{\square}$ of 
    $4n+1$ vertices, and a set $V_{u_{2}}^{\square}$ of $4n$ vertices. 
    Make $u_{1}$  adjacent to 
    every vertex of $\{u\}\cup V_{u}^{\triangle} \cup V_{u_{1}}^{\square}$, and make 
    $u_2$ adjacent to  every vertex of $V_{u}^{\triangle} \cup V_{u_{2}}^{\square}$.

     \item We introduce two new vertices $a$ and $b$ into $G'$, a set 
     $V_{a}^{\square}$ of  $|T|+\sum\limits_{u\in T} (d_{S}(u)-1) +2$ vertices, and 
     a set $V_{b}^{\square}$ of $|T|+1$  vertices. Make $a$ adjacent to 
     every vertex of $T\cup V_{a}^{\square} $, and make $b$ adjacent to very vertex of 
     $T\cup V_{b}^{\square}\cup\{x'\}$.
    Let $d_{S}(u)$, $u\in T$, denote the number  of 
neighbours of $u$ in $S$. For each $u\in T$, we add a set $V_{u}^{a}=\{u_{1}^{a},u_{2}^{a},\ldots,u_{d_{S}(u)-1}^{a}\}$ of $d_{S}(u)-1$ vertices
and make  $a$ and $u$ adjacent to every vertex of $V_{u}^{a}$. 
\item Finally, we add a set 
$T^{\square}=T_{1}^{\square} \cup T_{2}^{\square}$ of vertices where $|T_{1}^{\square}|=|T|+1$ and $|T_{2}^{\square}|=2$.
We make every vertex of $ V_{x}^{\triangle} \cup \bigcup\limits_{u\in T} V_{u}^{\triangle}$ adjacent to every vertex of $T_{2}^{\square}$. For every  $s\in S$, we make $s$ adjacent to $d_{T}(s)+1$ many arbitrary vertices of $T^{\square}_1$. For every  $t\in T^{\square}$, we 
introduce $d+2$  vertices and make them adjacent to  $t$ where $d$ is the degree of $t$ until this point of the construction; {\it these vertices are not shown in the figure}.  Finally, we set $k'=(|S|-k)+4n+2+|T|(4n+3)+1$.
\end{itemize}
\noindent Now we claim that there exists a vertex set $X \subseteq S$ 
in $G$ of size at most $k$ such that every vertex in $T$ has at least one neighbour in $X$ if and only if 
$G'$ has a globally minimal defensive alliance of size at least $k'$. 
Suppose there is a vertex set $X \subseteq S$ in $G$ of size at most $k$ such 
that every vertex in $T$ has at least one neighbour in $X$. 
We show that the set 
$$H= (S\setminus X)\cup \{b,x,x'\}\cup V_{x}^{\triangle} \cup \bigcup\limits_{u\in T} (\{u,u_{1},u_{2}\} \cup V_{u}^{\triangle})\cup \bigcup\limits_{u\in T} \{u_{1}^{a},\ldots,u_{d_{X}(u)-1}^{a}\}
$$ is a globally minimal defensive alliance. Clearly  $|H|\geq k'$ and observe that every vertex in $H$ is marginally protected. 
Let $u$ be an arbitrary element of $H$. If $u$ is an element of $T$, then $N_H(u)=\{b,u_1\}\cup\{u_{1}^{a},\ldots,u_{d_{X}(u)-1}^{a}\}\cup N_{S\setminus X}(u)$ and 
$N_{H^c}(u)=\{b\} \cup \{u_{d_X(u)},\ldots, u_{d_S(u)-1}\}\cup N_X(u)$. That is, $d_H(u)=d_S(u)+1$ and $d_{H^c}(u)=d_S(u)+1$;  thus $u$ is marginally protected. Similarly, it is easy to check that other vertices of $H$ are also marginally protected.  
We also see that  $G'[H]$ is connected. Using Observation 2,  $H$ is a globally minimal defensive alliance. 
\par To prove the reverse direction of the equivalence, we assume that there exists a globally minimal defensive alliance $H$ of size at least $k'$. By Observation \ref{onedegree}, $H$ can never contain a vertex of degree one.  As  one degree vertices 
cannot be part of the solution, we observe that no vertex from $T^{\square}$ is part of the solution. This is true because we will not be able to protect any vertex from $T^{\square}$ as more than half of the neighbours are one degree vertices. We claim that $T \subseteq H$. For the sake of contradiction suppose that there exists some $u\in T$ such that $u\not\in H$. Then vertices in $V_{u}^{\triangle} \cup \{u_{1},u_{2}\}$ cannot be inside $H$ as globally minimal defensive alliance must be connected. If we do not include $V_{u}^{\triangle}$ inside $H$ then $H$ cannot achieve the size $k'$. Therefore, we must include $u$ in $H$. From the above argument, we also see that we must include  $u_1$ and $u_2$ inside $H$ as otherwise vertices in $V_{u}^{\triangle}$ cannot be protected. Therefore, we have $\bigcup\limits_{u\in T} (V_{u}^{\triangle} \cup \{u,u_{1},u_{2}\}) \subseteq H$. Similarly, we can argue that $V_{x}^{\triangle} \cup \{x,x',b\} \subseteq H$. Observe that $u\in T$ must be marginally protected in $H$ as otherwise $H\setminus (V_{u}^{\triangle} \cup \{u_{1},u_{2}\})$ forms a defensive alliance, which is not possible. For $u\in T$, we have $N(u)= N_{S}(u) \cup V_{u}^{a} \cup \{a,b,u_{1}\}$
and hence $d(u) = 2d_{S}(u)+2$. Since $a\not\in H$ and $b,u_{1}\in H$, we must have added at most $d_{S}(u)-1$ nodes from  $N_S(u)$ in $H$ for each $u\in T$. Consider $X=H^{c}\cap S$. Clearly, every vertex $u\in T$ has at least one neighbour in $X$. Next, we see that $|X|\leq k$. As otherwise,  $|S\cap H|< |S|-k$ and then  $x$ cannot be protected. This implies that $I$ is a yes instance. \qed

\section{Hardness Results} 
 In this section,  we show that  the {\sc Globally Minimal Defensive Alliance} problem is W[1]-hard when parameterized by  
the size of a vertex deletion set into trees of height at most three, i.e., 
a subset $R$ of the vertices of the graph such that every component in the graph, after removing $R$, is a tree of height at most three. We give a reduction 
from  the {\sc Multi-Colored Clique} problem. The input of {\sc Multi-Colored Clique} consists of a graph $G$, an integer $k$, and a partition $(V_1,\ldots,V_k)$ of the vertices of $G$; the task is to decide if there is a $k$-clique containing exactly one vertex from each set $V_i$.

 \begin{theorem}\label{treetheorem}\rm
 The {\sc Globally Minimal Defensive Alliance} problem is W[1]-hard when parameterized by  the size of a vertex deletion set into 
 trees of height at most 3.

 \end{theorem}
 
\proof  The approach for using {\sc Multi-Colored Clique} in reductions is described in \cite{FELLOWS200953}, and has been proven to be very useful in
showing hardness results in the parameterized complexity setting. 
We use $G$ to denote a graph colored with $k$ colors given in  an instance of {\sc Multi-Colored
Clique}, and $G'$ to denote the graph in the reduced instance of {\sc Globally Minimal Defensive Alliance}. For a color $i \in [k]$, let $V_i$ denote the
subset of vertices in $G$ colored with color $i$ and for a pair of distinct colors $i, j \in [k]$, let $E_{ij}$ denote the subset of
edges in $G$ with endpoints colored $i$ and $j$. 

\par We construct $G'$ using two types of gadgets. Our goal is to guarantee that any globally minimal defensive alliance in $G'$ with a specific size
encodes a multi-colored clique in $G$. These gadgets are the selection and validation gadgets. The selection gadgets encode the selection of $k$ vertices and $k\choose 2$
edges that together encode a vertex and edge set of some multi-colored clique in $G$. The selection gadgets also ensure that in fact $k$ distinct vertices are chosen from $k$ distinct color classes, and that distinct $k\choose 2$
edges are chosen from $k\choose 2$ distinct edge color classes. The validation gadgets validate the selection done in the selection gadgets in the sense that they make sure that the edges chosen are in fact incident to the selected vertices. In the following
we sketch the construction of selection and validation gadgets as given in \cite{BENZWI201187}:

\noindent{\it Selection:} For each color-class $i\in [k]$, and each pair of distinct colors $i,j \in [k]$, we construct a $i$-selection gadget and
a $\{i,j\}$-selection gadget which respectively encode the selection of a vertex colored $i$ and an edge colored $\{i,j\}$ in $G$.
The $i$-selection gadget consists of a vertex $x_v$ for every vertex $v \in V_i$ , and likewise, the $\{i,j\}$-selection gadget consists
of a vertex $x_{\{u,v\}}$ for every edge $\{u, v\} \in E_{\{i,j\}}$. There are no edges between the vertices of the selection gadgets, that is, the union of all vertices in these gadgets is an independent set in $G'$.\\

\noindent{\it Validation:} We assign to every vertex $v$ in $G$ two unique identification numbers, $\text{low}(v)$ and $\text{high}(v)$, with $\text{low}(v) \in [n]$
and $\text{high}(v) = 2n - \text{low}(v)$. For every pair of distinct colors $i,j \in [k]$, we construct validation gadgets between the
$\{i,j\}$-selection gadget and the $i$- and $j$-selection gadget. Let $i$ and $j$ be any pair of distinct colors. We describe
the validation gadget between the $i$- and $\{i,j\}$-selection gadgets. 
It consists of two validation vertices $\alpha_{ij}$ and $\beta_{ij}$, the 
{\it validation-pair} of this gadget. The first vertex $\alpha_{ij}$ of 
this pair is adjacent to each vertex $x_u, u \in V_{i}$, by $\text{high}(u)$ parallel edges, and to each
edge-selection vertex $x_{\{u,v\}}, \{u, v\} \in E_{\{ij\}}$ and $v \in V_{j}$ , by $\text{low}(u)$ parallel edges. The other vertex $\beta_{ij}$ is adjacent to
each $x_u, u \in V_{i}$, by $\text{low}(u)$ parallel edges, and to each $x_{\{u,v\}}, \{u, v\} \in E_{\{ij\}}$ and $v \in V_{j}$, by $\text{high}(u)$ parallel edges. We next subdivide the edges between the selection and validation gadgets to obtain a simple graph, where all new vertices introduced by the subdivision are referred to as the {\it connection vertices}. The set of connection vertices adjacent to one of the
validation vertices
$\{\alpha_{ij}, \beta_{ij}\}$
and $x_{u}$ is denoted by $A_{ij}^u$ and  the set of connection vertices 
adjacent to one of the validation vertices $\{\alpha_{ij},\beta_{ij}\}$ and $x_{\{u,v\}}$ is denoted by $B_{ij}^{uv}$.
\par For each connection vertex  we add two new vertices and make them adjacent to 
it. 
For each $x_u$ in $i$-selection gadget,  we introduce $d(x_u)$ new vertices and make them
adjacent to  $x_u$, and likewise for   each $x_{\{u,v\}}$ in $\{i,j\}$-selection gadget we introduce $d(x_{\{u,v\}})$ new vertices and make them
adjacent to $x_{\{u,v\}}$. Let $L$
be the set of all validation vertices in $G'$. Set $N=100n^2$. \\

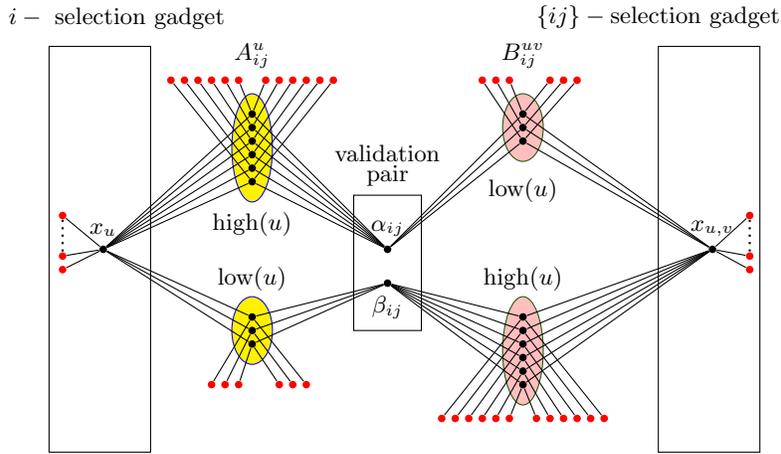
\begin{figure}
    \centering
\begin{tikzpicture}[scale=0.9]
\draw[black] (0,0) rectangle (1.5,6);

\draw[black] (4.5,1.8) rectangle (5.5,3.8);

\draw[black] (9,0) rectangle (10.5,6);
\draw[blue, fill=yellow] (3,4.5) ellipse (0.3cm and .8cm);
\draw[blue, fill=yellow] (3,1.8) ellipse (0.3cm and .5cm);

\draw[green, fill=pink] (7,4.8) ellipse (0.3cm and .5cm);
\draw[green, fill=pink] (7,1.5) ellipse (0.3cm and .8cm);
\node[circle, fill=black, inner sep=0 pt, minimum size=0.1cm] (l1) at (3,2) []{};
\node[circle, fill=black, inner sep=0 pt, minimum size=0.1cm] (l2) at (3,1.8) []{};
\node[circle, fill=black, inner sep=0 pt, minimum size=0.1cm] (l3) at (3,1.6) []{};

\node[circle, fill=black, inner sep=0 pt, minimum size=0.1cm] (l4) at (3,5) []{};
\node[circle, fill=black, inner sep=0 pt, minimum size=0.1cm] (l5) at (3,4.8) []{};
\node[circle, fill=black, inner sep=0 pt, minimum size=0.1cm] (l6) at (3,4.6) []{};
\node[circle, fill=black, inner sep=0 pt, minimum size=0.1cm] (l7) at (3,4.4) []{};
\node[circle, fill=black, inner sep=0 pt, minimum size=0.1cm] (l8) at (3,4.2) []{};
\node[circle, fill=black, inner sep=0 pt, minimum size=0.1cm] (l9) at (3,4) []{};

\node[circle, fill=black, inner sep=0 pt, minimum size=0.1cm] (r1) at (7,5) []{};
\node[circle, fill=black, inner sep=0 pt, minimum size=0.1cm] (r2) at (7,4.8) []{};
\node[circle, fill=black, inner sep=0 pt, minimum size=0.1cm] (r3) at (7,4.6) []{};

\node[circle, fill=black, inner sep=0 pt, minimum size=0.1cm] (r4) at (7,2) []{};
\node[circle, fill=black, inner sep=0 pt, minimum size=0.1cm] (r5) at (7,1.8) []{};
\node[circle, fill=black, inner sep=0 pt, minimum size=0.1cm] (r6) at (7,1.6) []{};
\node[circle, fill=black, inner sep=0 pt, minimum size=0.1cm] (r7) at (7,1.4) []{};
\node[circle, fill=black, inner sep=0 pt, minimum size=0.1cm] (r8) at (7,1.2) []{};
\node[circle, fill=black, inner sep=0 pt, minimum size=0.1cm] (r9) at (7,1) []{};

\node[circle, fill=black, inner sep=0 pt, minimum size=0.1cm] (xu) at (.8,3) [label=above:$x_{u}$]{};

\node[circle, fill=red, inner sep=0 pt, minimum size=0.1cm] (x1) at (.2,2.7) []{};
\node[circle, fill=red, inner sep=0 pt, minimum size=0.1cm] (x2) at (.2,2.9) []{};
\node[circle, fill=red, inner sep=0 pt, minimum size=0.1cm] (x3) at (.2,3.5) []{};

\draw[thick, dotted](x3)--(x2);

\draw(xu)--(x1);
\draw(xu)--(x2);
\draw(xu)--(x3);

\node[circle, fill=red, inner sep=0 pt, minimum size=0.1cm] (y1) at (10.35,2.7) []{};
\node[circle, fill=red, inner sep=0 pt, minimum size=0.1cm] (y2) at (10.35,2.9) []{};
\node[circle, fill=red, inner sep=0 pt, minimum size=0.1cm] (y3) at (10.35,3.5) []{};

\draw[thick, dotted](y3)--(y2);

\node[circle, fill=black, inner sep=0 pt, minimum size=0.1cm] (xuv) at (9.8,3) [label=above:$x_{u,v}$]{};

\node[circle, fill=black, inner sep=0 pt, minimum size=0.1cm] (m1) at (5,3) [label=above:$\alpha_{ij}$]{};
\node[circle, fill=black, inner sep=0 pt, minimum size=0.1cm] (m2) at (5,2.5) [label=below:$\beta_{ij}$]{};

\draw(xuv)--(y1);
\draw(xuv)--(y2);
\draw(xuv)--(y3);

\draw(xu)--(l1);
\draw(xu)--(l2);
\draw(xu)--(l3);
\draw(xu)--(l4);
\draw(xu)--(l5);
\draw(xu)--(l6);
\draw(xu)--(l7);
\draw(xu)--(l8);
\draw(xu)--(l9);

\draw(xuv)--(r1);
\draw(xuv)--(r2);
\draw(xuv)--(r3);
\draw(xuv)--(r4);
\draw(xuv)--(r5);
\draw(xuv)--(r6);
\draw(xuv)--(r7);
\draw(xuv)--(r8);
\draw(xuv)--(r9);

\draw(m1)--(l4);
\draw(m1)--(l5);
\draw(m1)--(l6);
\draw(m1)--(l7);
\draw(m1)--(l8);
\draw(m1)--(l9);

\draw(m1)--(r1);
\draw(m1)--(r2);
\draw(m1)--(r3);

\draw(m2)--(r4);
\draw(m2)--(r5);
\draw(m2)--(r6);
\draw(m2)--(r7);
\draw(m2)--(r8);
\draw(m2)--(r9);

\draw(m2)--(l1);
\draw(m2)--(l2);
\draw(m2)--(l3);

\node[circle, fill=red, inner sep=0 pt, minimum size=0.1cm] (f1) at (2.8,5.5) []{};
\node[circle, fill=red, inner sep=0 pt, minimum size=0.1cm] (f2) at (2.6,5.5) []{};
\node[circle, fill=red, inner sep=0 pt, minimum size=0.1cm] (f3) at (2.4,5.5) []{};
\node[circle, fill=red, inner sep=0 pt, minimum size=0.1cm] (f4) at (2.2,5.5) []{};
\node[circle, fill=red, inner sep=0 pt, minimum size=0.1cm] (f5) at (2.0,5.5) []{};
\node[circle, fill=red, inner sep=0 pt, minimum size=0.1cm] (f6) at (1.8,5.5) []{};

\node[circle, fill=red, inner sep=0 pt, minimum size=0.1cm] (f7) at (3.2,5.5) []{};
\node[circle, fill=red, inner sep=0 pt, minimum size=0.1cm] (f8) at (3.4,5.5) []{};
\node[circle, fill=red, inner sep=0 pt, minimum size=0.1cm] (f9) at (3.6,5.5) []{};
\node[circle, fill=red, inner sep=0 pt, minimum size=0.1cm] (f10) at (3.8,5.5) []{};
\node[circle, fill=red, inner sep=0 pt, minimum size=0.1cm] (f11) at (4.0,5.5) []{};
\node[circle, fill=red, inner sep=0 pt, minimum size=0.1cm] (f12) at (4.2,5.5) []{};

\node[circle, fill=red, inner sep=0 pt, minimum size=0.1cm] (f13) at (6.8,.5) []{};
\node[circle, fill=red, inner sep=0 pt, minimum size=0.1cm] (f14) at (6.6,.5) []{};
\node[circle, fill=red, inner sep=0 pt, minimum size=0.1cm] (f15) at (6.4,.5) []{};
\node[circle, fill=red, inner sep=0 pt, minimum size=0.1cm] (f16) at (6.2,.5) []{};
\node[circle, fill=red, inner sep=0 pt, minimum size=0.1cm] (f17) at (6.0,.5) []{};
\node[circle, fill=red, inner sep=0 pt, minimum size=0.1cm] (f18) at (5.8,.5) []{};

\node[circle, fill=red, inner sep=0 pt, minimum size=0.1cm] (f19) at (7.2,.5) []{};
\node[circle, fill=red, inner sep=0 pt, minimum size=0.1cm] (f20) at (7.4,.5) []{};
\node[circle, fill=red, inner sep=0 pt, minimum size=0.1cm] (f21) at (7.6,.5) []{};
\node[circle, fill=red, inner sep=0 pt, minimum size=0.1cm] (f22) at (7.8,.5) []{};
\node[circle, fill=red, inner sep=0 pt, minimum size=0.1cm] (f23) at (8.0,.5) []{};
\node[circle, fill=red, inner sep=0 pt, minimum size=0.1cm] (f24) at (8.2,.5) []{};

\node[circle, fill=red, inner sep=0 pt, minimum size=0.1cm] (f25) at (6.8,5.5) []{};
\node[circle, fill=red, inner sep=0 pt, minimum size=0.1cm] (f26) at (6.6,5.5) []{};
\node[circle, fill=red, inner sep=0 pt, minimum size=0.1cm] (f27) at (6.4,5.5) []{};
\node[circle, fill=red, inner sep=0 pt, minimum size=0.1cm] (f28) at (7.4,5.5) []{};
\node[circle, fill=red, inner sep=0 pt, minimum size=0.1cm] (f29) at (7.6,5.5) []{};
\node[circle, fill=red, inner sep=0 pt, minimum size=0.1cm] (f30) at (7.8,5.5) []{};

\node[circle, fill=red, inner sep=0 pt, minimum size=0.1cm] (f31) at (2.8,1) []{};
\node[circle, fill=red, inner sep=0 pt, minimum size=0.1cm] (f32) at (2.6,1) []{};
\node[circle, fill=red, inner sep=0 pt, minimum size=0.1cm] (f33) at (2.4,1) []{};
\node[circle, fill=red, inner sep=0 pt, minimum size=0.1cm] (f34) at (3.4,1) []{};
\node[circle, fill=red, inner sep=0 pt, minimum size=0.1cm] (f35) at (3.6,1) []{};
\node[circle, fill=red, inner sep=0 pt, minimum size=0.1cm] (f36) at (3.8,1) []{};

\draw(l4)--(f1);
\draw(l5)--(f2);
\draw(l6)--(f3);
\draw(l7)--(f4);
\draw(l8)--(f5);
\draw(l9)--(f6);

\draw(l4)--(f7);
\draw(l5)--(f8);
\draw(l6)--(f9);
\draw(l7)--(f10);
\draw(l8)--(f11);
\draw(l9)--(f12);

\draw(r4)--(f18);
\draw(r5)--(f17);
\draw(r6)--(f16);
\draw(r7)--(f15);
\draw(r8)--(f14);
\draw(r9)--(f13);

\draw(r4)--(f24);
\draw(r5)--(f23);
\draw(r6)--(f22);
\draw(r7)--(f21);
\draw(r8)--(f20);
\draw(r9)--(f19);

\draw(r1)--(f25);
\draw(r2)--(f26);
\draw(r3)--(f27);

\draw(r1)--(f28);
\draw(r2)--(f29);
\draw(r3)--(f30);

\draw(l1)--(f36);
\draw(l2)--(f35);
\draw(l3)--(f34);

\draw(l1)--(f33);
\draw(l2)--(f32);
\draw(l3)--(f31);

\node at (3,6.3) [label=below:$A_{ij}^u$]{};
\node at (7,6.3) [label=below:$B_{ij}^{uv}$]{};
\node at (3,3.8) [label=below:high$(u)$]{};

\node at (3,3) [label=below:low$(u)$]{};

\node at (7,4.3) [label=below:low$(u)$]{};

\node at (7,3) [label=below:high$(u)$]{};

\node at (5,4.8) [label=below:$\text{validation}$]{};
\node at (5,4.5) [label=below:$\text{pair}$]{};

\node at (1,6) [label=above:$i- \ \text{selection gadget}$]{};

\node at (9,6) [label=above:$\{ij\} - \text{selection gadget}$]{};

\end{tikzpicture}
    \caption{A graphical depiction of the validation gadget. In the example, $n = 5$ and low($u$) = 3. Note that $A^{u}_{ij}$ contains the connection vertices in yellow region and 
    $B^{uv}_{ij}$ contains the connection vertices in pink region. The red vertices are
    one degree vertices, and hence they are not part of any globally minimal 
    defensive alliance.}
    \label{fig:my_label}
\end{figure}
For every validation vertex  $\alpha\in L$, we add the 
following {\it protection gadget}. We introduce a new vertex $\alpha^{\triangle}$, a set 
$V^{\triangle}_{\alpha}$ of $N$ vertices, and a set of $V^{\square}_{\alpha^{\triangle}}$ of $N$ vertices. We make 
$\alpha$ adjacent to every vertex of $V^{\triangle}_{\alpha}$; make 
$\alpha^{\triangle}$ adjacent to every vertex of 
$V^{\triangle}_{\alpha} \cup V^{\square}_{\alpha^{\triangle}}$.
For each vertex $x\in V^{\triangle}_{\alpha}$, we introduce two new vertices
and make them adjacent to $x$. \\

\begin{figure}
    \centering
 \begin{tikzpicture}[scale=0.8]
 
\node[circle, fill=black, inner sep=0 pt, minimum size=0.1cm] (x) at (0,0) [label=left:$\alpha$]{};

\node[circle, fill=black, inner sep=0 pt, minimum size=0.1cm] (x2) at (0.6,1.5) []{};
\node[circle, fill=black, inner sep=0 pt, minimum size=0.1cm] (x3) at (-0.4,1.5) []{};
\node[circle, fill=black, inner sep=0 pt, minimum size=0.1cm] (x4) at (-0.7,1.5) []{};

\node[circle, fill=red, inner sep=0 pt, minimum size=0.1cm] (x8) at (0.6,4.5) []{};
\node[circle, fill=red, inner sep=0 pt, minimum size=0.1cm] (x9) at (-0.4,4.5) []{};
\node[circle, fill=red, inner sep=0 pt, minimum size=0.1cm] (x10) at (-0.6,4.5) []{};

\node[circle, fill=red, inner sep=0 pt, minimum size=0.1cm] (n1) at (0.9,1.7) []{};
\node[circle, fill=red, inner sep=0 pt, minimum size=0.1cm] (n2) at (0.9,2.2) []{};
\node[circle, fill=red, inner sep=0 pt, minimum size=0.1cm] (n3) at (0.9,2.4) []{};

\node[circle, fill=red, inner sep=0 pt, minimum size=0.1cm] (n4) at (0.9,1.3) []{};
\node[circle, fill=red, inner sep=0 pt, minimum size=0.1cm] (n5) at (0.9,0.8) []{};
\node[circle, fill=red, inner sep=0 pt, minimum size=0.1cm] (n6) at (0.9,0.6) []{};

\draw(x2)--(n1);
\draw(x3)--(n2);
\draw(x4)--(n3);
\draw(x2)--(n4);
\draw(x3)--(n5);
\draw(x4)--(n6);

\draw[dotted,thick](n1)--(n2);
\draw[dotted,thick](n4)--(n5);

\draw[dotted,thick](x8)--(x9);

\draw[dotted,thick](x3)--(x2);

\node[circle, fill=black, inner sep=0 pt, minimum size=0.1cm] (x0) at (0,3) [label=left:$\alpha^{\triangle}$]{};

\node[circle, fill=red, inner sep=0 pt, minimum size=0.1cm] (x5) at (0.6,-1.5) []{};
\node[circle, fill=red, inner sep=0 pt, minimum size=0.1cm] (x6) at (-0.4,-1.5) []{};
\node[circle, fill=red, inner sep=0 pt, minimum size=0.1cm] (x7) at (-0.6,-1.5) []{};

\draw[dotted,thick](x6)--(x5);

\node[circle, fill=black, inner sep=0 pt, minimum size=0.1cm] (y) at (4,1.5) [label=right:$\alpha'$]{};

\node[circle, fill=black, inner sep=0 pt, minimum size=0.1cm] (y2) at (4.6,3) []{};
\node[circle, fill=black, inner sep=0 pt, minimum size=0.1cm] (y3) at (3.6,3) []{};
\node[circle, fill=black, inner sep=0 pt, minimum size=0.1cm] (y4) at (3.3,3) []{};

\node[circle, fill=red, inner sep=0 pt, minimum size=0.1cm] (c1) at (4.9,3.2) []{};
\node[circle, fill=red, inner sep=0 pt, minimum size=0.1cm] (c2) at (4.9,3.7) []{};
\node[circle, fill=red, inner sep=0 pt, minimum size=0.1cm] (c3) at (4.9,3.9) []{};

\node[circle, fill=red, inner sep=0 pt, minimum size=0.1cm] (c4) at (4.9,2.8) []{};
\node[circle, fill=red, inner sep=0 pt, minimum size=0.1cm] (c5) at (4.9,2.3) []{};
\node[circle, fill=red, inner sep=0 pt, minimum size=0.1cm] (c6) at (4.9,2.1) []{};

\draw(y2)--(c1);
\draw(y3)--(c2);
\draw(y4)--(c3);
\draw(y2)--(c4);
\draw(y3)--(c5);
\draw(y4)--(c6);

\draw[dotted,thick](y3)--(y2);

\node[circle, fill=black, inner sep=0 pt, minimum size=0.1cm] (y0) at (4,4.5) [label=right:$\alpha^{\prime \triangle}$]{};

\node[circle, fill=red, inner sep=0 pt, minimum size=0.1cm] (y5) at (4.6,0) []{};
\node[circle, fill=red, inner sep=0 pt, minimum size=0.1cm] (y6) at (3.3,-0) []{};
\node[circle, fill=red, inner sep=0 pt, minimum size=0.1cm] (y7) at (3.6,0) []{};

\node[circle, fill=red, inner sep=0 pt, minimum size=0.1cm] (y8) at (4.6,6) []{};
\node[circle, fill=red, inner sep=0 pt, minimum size=0.1cm] (y9) at (3.3,6) []{};
\node[circle, fill=red, inner sep=0 pt, minimum size=0.1cm] (y10) at (3.6,6) []{};

\draw[dotted,thick](y5)--(y7);
\draw[dotted,thick](y8)--(y10);
\draw(x)--(y);

\draw(x)--(x2);
\draw(x)--(x3);
\draw(x)--(x4);

\draw(x)--(x6);
\draw(x)--(x7);
\draw(x)--(x5);

\draw(x0)--(x2);
\draw(x0)--(x3);
\draw(x0)--(x4);

\draw(x0)--(x8);
\draw(x0)--(x9);
\draw(x0)--(x10);

\draw(y)--(y2);
\draw(y)--(y3);
\draw(y)--(y4);

\draw(y)--(y6);
\draw(y)--(y7);
\draw(y)--(y5);

\draw(y0)--(y2);
\draw(y0)--(y3);
\draw(y0)--(y4);

\draw(y0)--(y8);
\draw(y0)--(y9);
\draw(y0)--(y10);

\node at (0,4.5) [label=above:$V_{\alpha^{\triangle}}^{\square}$]{};
\node at (-0.5,1.5) [label=left:$V_{\alpha}^{\triangle}$]{};

\node at (0,-1.5) [label=below:$V_{\alpha}^{\square}$]{};

\node at (4,6) [label=above:$V_{\alpha^{\prime \triangle}}^{\square}$]{};

\node at (3.3,3) [label=left:$V_{\alpha^{\prime}}^{\triangle}$]{};

\node at (4,0) [label=below:$V_{\alpha^{\prime}}^{\square}$]{};

 \end{tikzpicture}
    \caption{A graphical depiction of protection and marginal protection gadgets
    associated with validation vertex $\alpha$. Note that the red vertices are
    one degree vertices, and hence they are not part of any globally minimal 
    defensive alliance.}
    \label{fig:my_label1}
\end{figure}
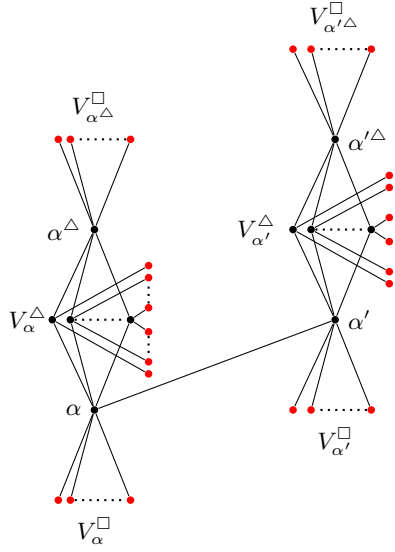

Finally for every validation vertex  
$\alpha\in L$, we add the 
following {\it marginal protection gadget}. This gadget ensures that 
$\alpha$ is marginally protected in globally minimal defensive alliance.
We introduce two new vertices $\alpha'$ and $\alpha'^{\triangle}$, 
a set $V^{\triangle}_{\alpha'}$ of $N$ vertices, a set $V^{\square}_{\alpha'}$
of $N$ vertices, and a set $V^{\square}_{\alpha'^{\triangle}}$ of $N+1$ vertices.
We make $\alpha'$ adjacent to every vertex of $\{\alpha\}\cup V^{\triangle}_{\alpha'} \cup V^{\square}_{\alpha'}$; make $\alpha'^{\triangle}$ adjacent to every vertex of  $V^{\triangle}_{\alpha'} \cup V^{\square}_{\alpha'^{\triangle}}$. For each vertex $x\in V^{\triangle}_{\alpha'}$, we introduce two new vertices and make them adjacent to $x$.
Let $d$ be the degree of $\alpha$ up to this point of construction. 
Finally for each $\alpha\in L$, we introduce a set $V_\alpha^{\square}$ of 
$N+4n-d+1$ vertices and make them adjacent to $\alpha$.
 This completes the construction of graph $G'$.\\
 We set $k' =k+ {k\choose 2}+ 8n{k\choose 2}+ 2(4N+8){k\choose 2}$. We observe that on removing 
 \begin{align*}
  R &=\big\{ \alpha_{ij}, \alpha_{ji}, \alpha^{\triangle}_{ij},
 \alpha^{\triangle}_{ji}, \alpha'_{ij},\alpha'_{ji}, \alpha'^{\triangle}_{ij}, \alpha'^{\triangle}_{ji}
 ~:~i,j\in [k], i\neq j\big\}\\
 & \cup \big\{ \beta_{ij},\beta_{ji}, \beta^{\triangle}_{ij},\beta^{\triangle}_{ji} \beta'_{ij},\beta'_{ji}, \beta'^{\triangle}_{ij}, \beta'^{\triangle}_{ji} ~:~i,j\in [k], i\neq j \big\}
 \end{align*}
 from graph $G'$, we are left with trees of height at most $3$. Note that $|R|=16 {k\choose 2}$, a function of $k$ only.  We claim that
  $G$ has a $k$-clique with exactly one vertex from each $V_i$ if and only if
  $G'$ has a globally minimal defensive alliance of size at least $k'$.
  Suppose that $v_1\in V_1$, $v_2\in V_2$,\ldots, $v_k\in V_k$ is a $k$-clique
  in $G$. We show that
\begin{align*}
S &=\big\{x_{v_i}~:~i\in [k] \big\} \cup \big \{x_{\{v_{i},v_{j}\}}~:~ i,j\in [k], i\neq j\big \} \cup        
 \bigcup\limits_{i,j\in [k], i\neq j}{A_{ij}^{v_i}\cup B_{ij}^{v_iv_j}\cup A_{ji}^{v_j}\cup B_{ji}^{v_jv_i}}\\
& \bigcup\limits_{i,j\in [k], i\neq j}{\big\{\alpha_{ij},  \alpha^{\triangle}_{ij},
  \alpha'_{ij}, \alpha'^{\triangle}_{ij}, \beta_{ij}, \beta^{\triangle}_{ij},  \beta'_{ij},\beta'^{\triangle}_{ij}\big\}}\cup V^{\triangle}_{\alpha_{ij}}\cup 
  V^{\triangle}_{\alpha'_{ij}}\cup V^{\triangle}_{\beta_{ij}}\cup V^{\triangle}_{\beta'_{ij}}\\
  & \bigcup\limits_{i,j\in [k], i\neq j}{\big\{\alpha_{ji},  \alpha^{\triangle}_{ji},
  \alpha'_{ji}, \alpha'^{\triangle}_{ji}, \beta_{ji}, \beta^{\triangle}_{ji},  \beta'_{ji},\beta'^{\triangle}_{ji}\big\}}\cup V^{\triangle}_{\alpha_{ji}}\cup 
  V^{\triangle}_{\alpha'_{ji}}\cup V^{\triangle}_{\beta_{ji}}\cup V^{\triangle}_{\beta'_{ji}}\\
 \end{align*}
is a globally minimal defensive alliance. Clearly $|S|=k'$. 
To prove that $S$ is a globally minimal defensive alliance, we prove that $S$ is connected and every vertex in $S$ is marginally protected. 
It is easy to see that each  vertex in  $\big\{x_{v_i}~:~i\in [k] \big\} \cup \big \{x_{\{v_{i},v_{j}\}}~:~ i,j\in [k], i\neq j\big \} $ is marginally protected as all the connection vertices adjacent to it are inside the solution and the same number of one degree neighbours are outside the solution. It is also easy to see that every connection vertex in
 $ \bigcup\limits_{i,j\in [k], i\neq j}{A_{ij}^{v_i}\cup B_{ij}^{v_iv_j}\cup A_{ji}^{v_j}\cup B_{ji}^{v_jv_i}}$
 is marginally protected as it has two neighbours inside the solution and two neighbours outside the solution. Similarly, we observe that all the vertices in the set 
\begin{align*}
& \bigcup\limits_{i,j\in [k], i\neq j}{\big\{  \alpha^{\triangle}_{ij},
  \alpha'_{ij}, \alpha'^{\triangle}_{ij},  \beta^{\triangle}_{ij},  \beta'_{ij},\beta'^{\triangle}_{ij}\big\}}\cup V^{\triangle}_{\alpha_{ij}}\cup 
  V^{\triangle}_{\alpha'_{ij}}\cup V^{\triangle}_{\beta_{ij}}\cup V^{\triangle}_{\beta'_{ij}}\\
  & \bigcup\limits_{i,j\in [k], i\neq j}{\big\{  \alpha^{\triangle}_{ji},
  \alpha'_{ji}, \alpha'^{\triangle}_{ji},  \beta^{\triangle}_{ji},  \beta'_{ji},\beta'^{\triangle}_{ji}\big\}}\cup V^{\triangle}_{\alpha_{ji}}\cup 
  V^{\triangle}_{\alpha'_{ji}}\cup V^{\triangle}_{\beta_{ji}}\cup V^{\triangle}_{\beta'_{ji}}\\
 \end{align*}
are also marginally protected. Lastly, we prove that the vertices in the set 
$ \bigcup\limits_{i,j\in [k], i\neq j}\{\alpha_{ij}, \beta_{ij}, \alpha_{ji},\beta_{ji}\}$  are marginally protected.  Consider $\alpha_{ij}$; it has total $2n$ connection vertices neighbours inside the solution as $\text{high}(u)+\text{low}(u)=2n$. Since $V^{\triangle}_{\alpha_{ij}} \subseteq S$ and $V^{\square}_{\alpha_{ij}} \cap S = \emptyset$, note that, including itself,  $\alpha_{ij}$ has  $ N+2n + 1$ neighbours
in $S$   and $d_{S^{c}}(\alpha_{ij}) = (d-2n) + (N+4n-d+2) = N+ 2n+1$. Therefore  $\alpha_{ij}$ is marginally protected. It is easy to observe that $S$ is connected. This shows that $S$ is a globally minimal defensive alliance.
\par In the reverse direction, we assume that $G'$ admits a globally minimal defensive alliance $S$ of size at least $k'$. By Observation 1, no vertex of degree one can be part of a globally minimal defensive alliance of size $\geq 2$ as an one degree vertex itself forms a defensive alliance. We  claim that the vertices in
\begin{align*}
& \bigcup\limits_{i,j\in [k], i\neq j}{\big\{\alpha_{ij},  \alpha^{\triangle}_{ij},
  \alpha'_{ij}, \alpha'^{\triangle}_{ij}, \beta_{ij}, \beta^{\triangle}_{ij},  \beta'_{ij},\beta'^{\triangle}_{ij}\big\}}\cup V^{\triangle}_{\alpha_{ij}}\cup 
  V^{\triangle}_{\alpha'_{ij}}\cup V^{\triangle}_{\beta_{ij}}\cup V^{\triangle}_{\beta'_{ij}}\\
  & \bigcup\limits_{i,j\in [k], i\neq j}{\big\{\alpha_{ji},  \alpha^{\triangle}_{ji},
  \alpha'_{ji}, \alpha'^{\triangle}_{ji}, \beta_{ji}, \beta^{\triangle}_{ji},  \beta'_{ji},\beta'^{\triangle}_{ji}\big\}}\cup V^{\triangle}_{\alpha_{ji}}\cup 
  V^{\triangle}_{\alpha'_{ji}}\cup V^{\triangle}_{\beta_{ji}}\cup V^{\triangle}_{\beta'_{ji}}
 \end{align*} are always in $S$. 
Assume, for the sake of contradiction, that $\alpha_{ij}\notin S$.  Then we cannot include any vertex from  $V^{\triangle}_{\alpha_{ij}}$ in the solution.
 This is true because every globally minimal defensive alliance must be connected.
If we cannot include the set $V^{\triangle}_{\alpha_{ij}}$ then clearly $|S|\leq k'$, a 
contradiction.  Next, we observe that protection of $\alpha_{ij}$ requires at least one vertex from the set $V^{\triangle}_{\alpha_{ij}}$ inside the solution. As every vertex in $V^{\triangle}_{\alpha_{ij}}$ has two one degree neighbours, it implies that the protection of that vertex requires $\alpha^{\triangle}_{ij}$ inside the solution. Now, the protection of $\alpha^{\triangle}_{ij}$ forces the full set $V^{\triangle}_{\alpha_{ij}}$
 inside the solution as its one degree neighbours are always outside $S$. Similarly, we argue that $\alpha'_{ij}$ and  $V^{\triangle}_{\alpha'_{ij}}$  will be inside the solution.  This proves the claim. \\
Observe that the above set has size exactly equal to $2(4N+8){k\choose 2}$. 
We need to add at least $k+ {k\choose 2}+8n{k\choose 2}$ more vertices in $S$. 
We claim that
$$\bigcup\limits_{i,j\in [k], i\neq j}{A_{ij}^{v_i}\cup B_{ij}^{v_iv_j}\cup A_{ji}^{v_j}\cup B_{ji}^{v_jv_i}}\subseteq S.$$
Observe that $\alpha_{ij}$ must be marginally protected inside the solution 
as otherwise 
$S\setminus (\{\alpha'_{ij}, \alpha'^{\triangle}_{ij}\}\cup V^{\triangle}_{\alpha'_{ij}})$
 forms a defensive alliance. This is equivalent to say that the marginal protection of $\alpha_{ij}$ requires  $2n$ neighbours from connection vertices inside the solution. Similarly, the marginal protection of $\beta_{ij}$ requires exactly $2n$ neighbours from connection vertices inside the solution. Therefore, for each $i,j\in [k]$, $i\neq j$,
 we have ${A_{ij}^{v_i}\cup B_{ij}^{v_iv_j}\cup A_{ji}^{v_j}\cup B_{ji}^{v_jv_i}}\subseteq S$.\\
 
We now show that each vertex (resp. edge) selection gadget contributes exactly one vertex in $S$.
We first show  that every vertex  selection gadget contributes at most one vertex inside the solution. Suppose there exists a vertex selection gadget which contributes at least two vertices inside the solution. Without loss of generality, suppose $x_{u_{1}}$ and $x_{u_{2}}$  from $i$-selection gadget are inside the solution. It implies that the protection of $x_{u_{1}}$ and $x_{u_{2}}$ requires $A_{ij}^{u_1}$ and $A_{ij}^{u_2}$ 
respectively inside the solution. Note that either $\alpha_{ij}$ or $\beta_{ij}$ will have more than $2n$ connection vertex neighbours inside the solution as either $\text{high}(u_{1}) + \text{high}(u_{2}) >2n$ or $\text{low}(u_{1}) + \text{low}(u_{2}) >2n$. This is a contradiction as either $\alpha_{ij}$ or $\beta_{ij}$
will not be marginally protected inside $S$. We can argue similarly for edge selection gadget and other color classes as well. Next, we show that every vertex (resp. edge) selection gadget contributes at least one vertex to solution. For the sake of contradiction, assume that $i$-selection gadget does not contribute any vertex to the solution. In this case, it will not be possible to protect the validation vertices  $\alpha_{ij}$ and $\beta_{ij}$ for $j\neq i$,  because no connection vertex between $i$-selection gadget and the validation pair $\{\alpha_{ij},\beta_{ij} \}$ can be added to solution. This is true as the protection of connection vertices  require their neighbour in $i$-selection gadget to be part of the solution. As the edge selection gadget can contribute at most one vertex due to above argument, the vertices in the set $\{\alpha_{ij},\beta_{ij} \}$ will have $<2n$ neighbours from the set of connection vertices. This makes the protection of the validation vertices  $\alpha_{ij}$ and $\beta_{ij}$ impossible. Therefore each selection gadget contributes exactly one vertex inside the solution.\\
 Next, we claim that if $i$-selection gadget contributes $x_{v_i}$ and
$j$-selection gadget contributes $x_{v_j}$ then $\{i,j\}$-selection gadget must contribute $x_{\{v_i,v_j\}}$. Assume, for the sake of contradiction, that $\{i,j\}$-selection gadget  contributes $x_{\{v_k,v_j\}}$ where $v_k\neq v_i$. In this case, we have 
$A^{v_i}_{ij}\cup A^{v_j}_{ji}\cup B^{v_iv_k}_{ij}\cup B^{v_jv_k}_{ji}\subseteq S$. 
Note  that one of the vertices from the set $\{ \alpha_{ij}, \beta_{ij}\}$ is not protected because when $v_i \neq v_k$ either high$(v_i)$+low$(v_k) < 2n$ or low$(v_i)$+high$(v_k) < 2n$. This is a contradiction. Therefore, we proved that if $i$-selection gadget contributes $x_{v_i}$ and
$j$-selection gadget contributes $x_{v_j}$, then $\{i,j\}$-selection gadget must contribute $x_{\{v_i,v_j\}}$. It implies that the set $\{v_i\in G~|~x_{v_i} \in i\mbox{-selection gadget}, i\in [k]\}$ forms a multicolored clique in $G$.\qed

 Clearly trees of height at most three are trivially acyclic. 
 Moreover, it is easy to verify that such trees have 
 pathwidth 
 \cite{Kloks94} 
 and treedepth 
 \cite{Sparsity} 
 at most three, which implies:
 
\begin{theorem}\rm
 The {\sc Globally Minimal Defensive Alliance} problem 
 is W[1]-hard when parameterized by any of the following parameters:
 \begin{itemize}
     \item the feedback vertex set number,
     \item the treewidth of the input graph,
     \item the pathwidth and treedepth of the input graph.
 \end{itemize}
 
\end{theorem}

\section{NP-completeness}
Rooted Globally Minimal Defensive Alliance asks for a globally minimal defensive alliance $S$ that contains a specified vertex $r$ in a graph $G$. The vertex $r$ is called the root of $S$. A globally minimal defensive alliance of $G$ will not be of use for this problem unless the set contains $r$. We define the problem as follows:
    \vspace{3mm}
    \\
    \fbox
    {\begin{minipage}{33.7em}\label{FFVS1 }
       {\sc Rooted Globally Minimal Defensive Alliance}\\
        \noindent{\bf Input:} An undirected graph $G=(V,E)$, a vertex $r\in V$.
    
        \noindent{\bf Question:} Does there exist a globally minimal defensive alliance
        $S\subseteq V$, such that $r\in S$?
    \end{minipage} }
  \vspace{3mm} 

\noindent In this section, we prove the following theorem:

 \begin{theorem}\label{Stheorem}\rm The {\sc Rooted Globally Minimal Defensive Alliance} problem is NP-complete.
 \end{theorem}
 
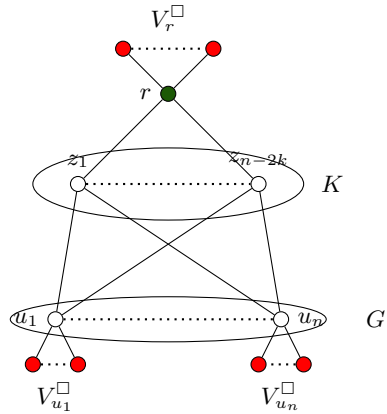
\begin{figure}[ht]
    \centering
\begin{tikzpicture}[scale=0.6]

	\node[circle,draw,fill=green, inner sep=0 pt, minimum size=0.2cm]	(x) at (0,0) [label=left:$r$]{};
	
	\node[circle,draw,fill=red, inner sep=0 pt, minimum size=0.2cm]	(x1) at (-1,1) []{};
	\node[circle,draw,fill=red, inner sep=0 pt, minimum size=0.2cm]	(x2) at (1,1) []{};
	
	\node(x0) at (0,1) [label=above:$V_{r}^{\square}$]{};

	\node[circle,draw, inner sep=0 pt, minimum size=0.2cm]	(y1) at (-2,-2) [label=above:$z_{1}$]{};
	\node[circle,draw, inner sep=0 pt, minimum size=0.2cm]	(y2) at (2,-2) [label=above:$z_{n-2k}$]{};
	\node[]() at (3,-2) [label=right:$K$]{};

\draw (0,-2) ellipse (3cm and 0.8cm);

\draw (0,-5) ellipse (3.5cm and 0.5cm);

\node[circle,draw, inner sep=0 pt, minimum size=0.2cm]	(u1) at (-2.5,-5) [label=left:$u_{1}$]{};
\node[circle,draw, inner sep=0 pt, minimum size=0.2cm]	(u2) at (2.5,-5) [label=right:$u_{n}$]{};
\node[]() at (4,-5) [label=right:$G$]{};

\draw[dotted, thick](y1)--(y2);
\draw[dotted, thick](u1)--(u2);




\node[circle,draw,fill=red, inner sep=0 pt, minimum size=0.2cm]	(u01) at (3,-6) []{};
\node[circle,draw,fill=red, inner sep=0 pt, minimum size=0.2cm]	(u02) at (2,-6) []{};

\node(x0) at (2.5,-6) [label=below:$V_{u_{n}}^{\square}$]{};
\node(x0) at (-2.5,-6) [label=below:$V_{u_{1}}^{\square}$]{};

\node[circle,draw,fill=red, inner sep=0 pt, minimum size=0.2cm]	(u11) at (-3,-6) []{};
\node[circle,draw,fill=red, inner sep=0 pt, minimum size=0.2cm]	(u12) at (-2,-6) []{};

\draw(x)--(x1);
\draw(x)--(x2);

\draw(u2)--(u01);
\draw(u2)--(u02);
\draw(u1)--(u11);
\draw(u1)--(u12);
\draw(x)--(y1);
\draw(x)--(y2);

\draw[dotted, thick](x1)--(x2);

\draw[dotted, thick](u01)--(u02);
\draw[dotted, thick](u11)--(u12);

\draw(y1)--(u1);
\draw(y1)--(u2);
\draw(y2)--(u1);
\draw(y2)--(u2);

\end{tikzpicture}
    \caption{Reduction from {\sc Clique} to {\sc Rooted Globally Minimal Defensive Alliance}}
    \label{fig:rooted}
\end{figure}

 \proof It is easy to see that  {\sc Rooted Globally Minimal Defensive Alliance} 
 is in NP. We prove it is NP-hard by giving a polynomial time reduction from {\sc Clique} on regular graphs to {\sc Rooted Globally Minimal Defensive Alliance}. See
 Figure \ref{fig:rooted} for an illustration. 
 Let $I=(G,k)$ be an instance of {\sc Clique}, where $G$ is an $s$-regular graph.  We construct an instance $I'=(G',r)$ of {\sc Rooted Globally Minimal Defensive Alliance}  as follows.
  First, we introduce one vertex $r$ and a set of vertices $\{z_1,z_1,\ldots,z_{n-2k}\}$ into $G'$. We make the set $\{z_1,z_1,\ldots,z_{n-2k}\}$ a clique $K$ in $G'$.  We make every vertex of $K$ adjacent  to every vertex $u$ of $G$ and the vertex $r$. Next, we introduce a set $V_{r}^{\square}$ of $n-2k$  vertices and make them adjacent to $r$.
 For every $u\in V(G)$, we introduce a set $V_{u}^{\square}$ of $n-s-2$  vertices 
 and make them adjacent to $u$. 
 This completes the construction of $G'$. To prove the correctness of the reduction, we claim that 
 $G$ has a $k$-clique if and only if $G'$ admits a globally minimal defensive alliance 
 containing $r$. Assume first that $G$ has a clique $C$ of size $k$. 
 We claim that $S = C \cup K \cup \{r\}$ is a globally minimal defensive alliance of $G'$. We observe that all the vertices in $S$ are marginally protected  and $G[S]$ is connected. Using Observation \ref{OBS2}, $S$ is a globally minimal defensive alliance containing $r$ in $G'$.  \\
 
 To prove the reverse direction of the equivalence, suppose $G'$ has a globally minimal defensive alliance $S$ containing vertex $r$. By Observation \ref{onedegree}, one degree vertices cannot be part of $S$. This implies that the protection of $r$ requires all the vertices of $K$ in the solution. Therefore, we can assume that $K \subseteq S$. We observe that every vertex of $K$ needs at least $k$ vertices from $V(G)$ for its protection. We also see that, if we take more than $k$ vertices from $V(G)$ inside $S$ then all the vertices in $K$ are overprotected. Then $S\setminus \{r\}$ is also a defensive alliance. This is a contradiction to the assumption that $S$ is globally minimal defensive alliance. This proves that $V(G)$ contributes exactly $k$ vertices in the solution. Let us denote this set by  $C$. Since $G$ is $s$-regular, note  that each $u\in C$ requires exactly $k-1$ neighbours from  $C$ for its protection. Then $u$ has $n-2k+k-1=n-k-1$ neighbours in $S$ and $(n-s-2)+(s-k+1)=n-k-1$ neighbours outside $S$. This implies that $C$ is a clique of size exactly $k$. \qed 

 \section{Conclusion} The main contributions in this paper are that the {\sc Globally Minimal Defensive Alliance} problem 
is FPT when parameterized by neighborhood diversity; 
no polynomial kernel parameterized by vertex cover number; and
 the problem  is W[1]-hard parameterized by 
a wide range of fairly restrictive structural parameters such as the feedback vertex set number, pathwidth,  treewidth, and  treedepth of the input graph. 
We also proved that given 
a vertex $v\in V(G)$, deciding if $G$ has a globally minimal defensive alliance
containing $v$, is NP-complete. 
 It would be interesting to consider the parameterized complexity with respect to  twin cover. The modular width parameter also appears to be a natural parameter to consider here; and since there are graphs with bounded modular-width and unbounded neighborhood diversity, we believe this is also an 
interesting open problem.  The parameterized complexity of the {\sc Globally Minimal Defensive Alliance} problem remains 
unsettled  when parameterized by the  solution size. 

\section*{Acknowledgement} The first author gratefully acknowledges support from the Ministry of Human Resource Development, 
 Government of India, under Prime Minister's Research Fellowship Scheme (No. MRF-192002-211).  
The second  author's research is supported in part by the Science and Engineering Research Board (SERB), Govt. of India, under Sanction Order No.
MTR/2018/001025.


\bibliographystyle{abbrv}
\bibliography{bibliography}

\begin{thebibliography}{10}

\bibitem{BAZGAN2019111}
C.~Bazgan, H.~Fernau, and Z.~Tuza.
\newblock Aspects of upper defensive alliances.
\newblock {\em Discrete Applied Mathematics}, 266:111 -- 120, 2019.

\bibitem{BENZWI201187}
O.~Ben-Zwi, D.~Hermelin, D.~Lokshtanov, and I.~Newman.
\newblock Treewidth governs the complexity of target set selection.
\newblock {\em Discrete Optimization}, 8(1):87--96, 2011.
\newblock Parameterized Complexity of Discrete Optimization.

\bibitem{BLIEM2018334}
B.~Bliem and S.~Woltran.
\newblock Defensive alliances in graphs of bounded treewidth.
\newblock {\em Discrete Applied Mathematics}, 251:334 -- 339, 2018.

\bibitem{Cami2006OnTC}
A.~Cami, H.~Balakrishnan, N.~Deo, and R.~Dutton.
\newblock On the complexity of finding optimal global alliances.
\newblock {\em J. Combin. Math. Combin. Comput.}, 58:23--31, 2006.

\bibitem{CHANG2012479}
C.-W. Chang, M.-L. Chia, C.-J. Hsu, D.~Kuo, L.-L. Lai, and F.-H. Wang.
\newblock Global defensive alliances of trees and cartesian product of paths
  and cycles.
\newblock {\em Discrete Applied Mathematics}, 160(4):479 -- 487, 2012.

\bibitem{marekcygan}
M.~Cygan, F.~V. Fomin, L.~Kowalik, D.~Lokshtanov, D.~Marx, M.~Pilipczuk,
  M.~Pilipczuk, and S.~Saurabh.
\newblock {\em Parameterized Algorithms}.
\newblock Springer, 2015.

\bibitem{Downey}
R.~G. Downey and M.~R. Fellows.
\newblock {\em Parameterized Complexity}.
\newblock Springer, 2012.

\bibitem{Enciso2009AlliancesIG}
R.~Enciso.
\newblock {\em Alliances in graphs: Parameterized algorithms and on
  partitioning series -parallel graphs}.
\newblock PhD thesis, USA, 2009.

\bibitem{FELLOWS200953}
M.~R. Fellows, D.~Hermelin, F.~Rosamond, and S.~Vialette.
\newblock On the parameterized complexity of multiple-interval graph problems.
\newblock {\em Theoretical Computer Science}, 410(1):53--61, 2009.

\bibitem{fellows}
M.~R. Fellows, D.~Lokshtanov, N.~Misra, F.~A. Rosamond, and S.~Saurabh.
\newblock Graph layout problems parameterized by vertex cover.
\newblock In S.-H. Hong, H.~Nagamochi, and T.~Fukunaga, editors, {\em
  Algorithms and Computation}, pages 294--305, Berlin, Heidelberg, 2008.
  Springer Berlin Heidelberg.

\bibitem{Fernau}
H.~Fernau and D.~Raible.
\newblock Alliances in graphs: a complexity-theoretic study.
\newblock In {\em Proceeding Volume II of the 33rd International Conference on
  Current Trends in Theory and Practice of Computer Science}, 2007.

\bibitem{10.5614/ejgta.2014.2.1.7}
H.~Fernau and J.~A. Rodriguez-Velazquez.
\newblock A survey on alliances and related parameters in graphs.
\newblock {\em Electronic Journal of Graph Theory and Applications}, 2(1),
  2014.

\bibitem{FERNAU2009177}
H.~Fernau, J.~A. Rodríguez, and J.~M. Sigarreta.
\newblock Offensive r-alliances in graphs.
\newblock {\em Discrete Applied Mathematics}, 157(1):177 -- 182, 2009.

\bibitem{fomin_lokshtanov_saurabh_zehavi_2019}
F.~V. Fomin, D.~Lokshtanov, S.~Saurabh, and M.~Zehavi.
\newblock {\em Kernelization: Theory of Parameterized Preprocessing}.
\newblock Cambridge University Press, 2019.

\bibitem{frick}
G.~Fricke, L.~Lawson, T.~Haynes, M.~Hedetniemi, and S.~Hedetniemi.
\newblock A note on defensive alliances in graphs.
\newblock {\em Bulletin of the Institute of Combinatorics and its
  Applications}, 38:37--41, 2003.

\bibitem{ICDCIT2021}
A.~Gaikwad, S.~Maity, and S.~K. Tripathi.
\newblock Parameterized complexity of defensive and offensive alliances in
  graphs.
\newblock In D.~Goswami and T.~A. Hoang, editors, {\em Distributed Computing
  and Internet Technology - 17th International Conference, {ICDCIT} 2021,
  January 7-10, 2021, Proceedings}, volume 12582 of {\em Lecture Notes in
  Computer Science}, pages 175--187. Springer, 2021.

\bibitem{HassanShafique2004PartitioningAG}
K.~Hassan-Shafique.
\newblock {\em Partitioning A Graph In Alliances And Its Application To Data
  Clustering}.
\newblock PhD thesis, USA, 2004.

\bibitem{Lindsay}
L.~H. Jamieson, S.~T. Hedetniemi, and A.~A. McRae.
\newblock The algorithmic complexity of alliances in graphs.
\newblock {\em Journal of Combinatorial Mathematics and Combinatorial
  Computing}, 68:137--150, 2009.

\bibitem{kannan}
R.~Kannan.
\newblock Minkowski's convex body theorem and integer programming.
\newblock {\em Mathematics of Operations Research}, 12(3):415--440, 1987.

\bibitem{Kloks94}
T.~Kloks.
\newblock {\em Treewidth, Computations and Approximations}, volume 842 of {\em
  Lecture Notes in Computer Science}.
\newblock Springer, 1994.

\bibitem{kris}
P.~Kristiansen, M.~Hedetniemi, and S.~Hedetniemi.
\newblock Alliances in graphs.
\newblock {\em Journal of Combinatorial Mathematics and Combinatorial
  Computing}, 48:157--177, 2004.

\bibitem{Lampis}
M.~Lampis.
\newblock Algorithmic meta-theorems for restrictions of treewidth.
\newblock {\em Algorithmica}, 64:19--37, 2012.

\bibitem{lenstra}
H.~W. Lenstra.
\newblock Integer programming with a fixed number of variables.
\newblock {\em Mathematics of Operations Research}, 8(4):538--548, 1983.

\bibitem{Manlove1998MinimaximalAM}
D.~Manlove.
\newblock {\em Minimaximal and maximinimal optimisation problems : a partial
  order-based approach}.
\newblock PhD thesis, Scotland, 1998.

\bibitem{Sparsity}
J.~Nesetril and P.~O. de~Mendez.
\newblock {\em Sparsity: Graphs, Structures, and Algorithms}.
\newblock Springer Publishing Company, Incorporated, 2014.

\bibitem{small}
C.~R., M.~M., R.~I., and S.~N.
\newblock Small alliances in graphs.
\newblock In {\em Kučera L., Kučera A. (eds) Mathematical Foundations of
  Computer Science, MFCS 2007, Lecture Notes in Computer Science}, volume 4708,
  Berlin, Heidelberg, 2007. Springer Berlin Heidelberg.

\bibitem{Neil}
N.~Robertson and P.~Seymour.
\newblock Graph minors. iii. planar tree-width.
\newblock {\em Journal of Combinatorial Theory, Series B}, 36(1):49 -- 64,
  1984.

\bibitem{ROD}
J.~Rodríguez-Velázquez and J.~Sigarreta.
\newblock Global offensive alliances in graphs.
\newblock {\em Electronic Notes in Discrete Mathematics}, 25:157 -- 164, 2006.

\bibitem{SIGARRETA20091687}
J.~Sigarreta, S.~Bermudo, and H.~Fernau.
\newblock On the complement graph and defensive k-alliances.
\newblock {\em Discrete Applied Mathematics}, 157(8):1687 -- 1695, 2009.

\bibitem{SIGARRETA20061345}
J.~Sigarreta and J.~Rodríguez.
\newblock On defensive alliances and line graphs.
\newblock {\em Applied Mathematics Letters}, 19(12):1345 -- 1350, 2006.

\bibitem{SIGA}
J.~Sigarreta and J.~Rodríguez.
\newblock On the global offensive alliance number of a graph.
\newblock {\em Discrete Applied Mathematics}, 157(2):219 -- 226, 2009.

\bibitem{Tedder}
M.~Tedder, D.~Corneil, M.~Habib, and C.~Paul.
\newblock Simpler linear-time modular decomposition via recursive factorizing
  permutations.
\newblock In {\em Automata, Languages and Programming}, pages 634--645, Berlin,
  Heidelberg, 2008. Springer Berlin Heidelberg.

\bibitem{west}
D.~B. West.
\newblock {\em Introduction to Graph Theory}.
\newblock Prentice Hall, 2000.

\end{thebibliography}



\end{document}